\documentclass[twocolumn,trackchanges,resetfootnote,twocolappendix]{aastex701}
\usepackage{amsmath}
\usepackage{xspace}

\newcommand{\Lbol}{\mbox{$L_{\rm bol}$}\xspace}
\newcommand{\Teff}{\mbox{$T_{\rm eff}$}\xspace}
\newcommand{\logg}{\mbox{$\log{g}$}\xspace}
\newcommand{\Msun}{\mbox{$M_{\sun}$}\xspace}
\newcommand{\Mjup}{\mbox{$M_{\rm Jup}$}\xspace}

\newcommand{\Rjup}{\mbox{$R_{\rm Jup}$}\xspace}

\newcommand{\Lp}{\mbox{$L^{\prime}$}\xspace}

\usepackage{xcolor}
\usepackage[dvipsnames]{xcolor} 

\newcommand{\eris}{VLT/ERIS\xspace}
\newcommand{\fours}{4S\xspace}

\DeclareRobustCommand{\asc_d}{$210.8$\raisebox{0.5ex}{\tiny$\substack{+0.6 \\ -0.4}$}} 
\DeclareRobustCommand{\inc_d}{$89.0$\raisebox{0.5ex}{\tiny$\substack{+0.7 \\ -0.6}$}} 
\DeclareRobustCommand{\sau_d}{$26.0$\raisebox{0.5ex}{\tiny$\substack{+2.2 \\ -6.1}$}} 
\DeclareRobustCommand{\per_d}{$91$\raisebox{0.5ex}{\tiny$\substack{+18 \\ -27}$}} 
\DeclareRobustCommand{\mass_d}{\mbox{$2.4\pm0.6$}}
\DeclareRobustCommand{\teff_d}{$600$\raisebox{0.5ex}{\tiny$\substack{+45 \\ -60}$}}
\DeclareRobustCommand{\radius_d}{\mbox{$1.26\pm0.03$}}
\DeclareRobustCommand{\age_myr}{\mbox{$23\pm8$}}

\shorttitle{Discovery of $\beta$~Pic~d from the ground \& space}
\shortauthors{B. J. Sutlieff \& M. J. Bonse et al.}
\received{May 26, 2026}
\revised{June 16, 2026}
\accepted{June 17, 2026}

\begin{document}

\title{Direct Imaging Discovery of Giant Exoplanet $\beta$~Pictoris~d: A Decade-Long Game of Hide-and-Seek}

\author[orcid=0000-0002-9962-132X,sname='Sutlieff']{Ben J. Sutlieff}
\altaffiliation{These two authors contributed equally to this work and should be considered joint first authors.}
\affiliation{Institute for Astronomy, University of Edinburgh, Royal Observatory, Blackford Hill, Edinburgh, EH9 3HJ, UK}
\affiliation{Centre for Exoplanet Science, University of Edinburgh, Edinburgh, EH9 3HJ, UK}
\email[show]{ben.sutlieff@roe.ac.uk}  

\author[orcid=0000-0003-2202-1745,sname='Bonse']{Markus J. Bonse} 
\altaffiliation{These two authors contributed equally to this work and should be considered joint first authors.}
\affiliation{European Southern Observatory, Karl-Schwarzschild-Stra\ss{}e 2, 85748 Garching bei M\"unchen, Germany}
\affiliation{Max Planck Institute for Intelligent Systems, Max-Planck-Ring 4, 72076 T\"ubingen, Germany}
\email[show]{markus.bonse@eso.org}

\author[orcid=0000-0002-0101-8814,sname='Christiaens']{Valentin Christiaens}
\affiliation{Universit\'e Paris-Saclay, Universit\'e Paris Cit\'e, CEA, CNRS, AIM, F-91191 Gif-sur-Yvette, France}
\affiliation{Universit\'e Paris-Saclay, CNRS, Institut d’Astrophysique Spatiale, 91405 Orsay, France}
\affiliation{STAR Institute, Universit\'e de Li\`ege, All\'ee du Six Ao\^ut 19c, 4000 Li\`ege, Belgium}
\email{valentin.christiaens@cea.fr}

\author[orcid=0000-0002-2428-9932,sname='Fontanive']{Cl\'emence Fontanive}
\affiliation{Institute for Astronomy, University of Edinburgh, Royal Observatory, Blackford Hill, Edinburgh, EH9 3HJ, UK}
\affiliation{Centre for Exoplanet Science, University of Edinburgh, Edinburgh, EH9 3HJ, UK}
\email{clemence.fontanive@roe.ac.uk}

\author[orcid=0000-0003-0593-1560,sname='Matthews']{Elisabeth C. Matthews}
\affiliation{Max-Planck-Institut f\"ur Astronomie, K\"onigstuhl 17, D-69117 Heidelberg, Germany}
\affiliation{Department of Astrophysics, American Museum of Natural History, Central Park West at 79th Street, New York, NY 10024, USA}
\email{matthews@mpia.de}

\author[orcid=0009-0009-8137-9991,sname='Parker']{Luke T. Parker}
\affiliation{Astrophysics, Department of Physics, University of Oxford, Denys Wilkinson Building, Keble Road, Oxford, OX1 3RH, UK}
\email{luke.parker@physics.ox.ac.uk}

\author[orcid=0000-0001-5653-5635,sname='Pearce']{Tim D. Pearce}
\affiliation{Department of Physics, University of Warwick, Gibbet Hill Road, Coventry CV4 7AL, UK}
\email{tim.pearce@warwick.ac.uk}

\author[orcid=0000-0002-4125-0140,sname='Birkby']{Jayne L. Birkby}
\affiliation{Astrophysics, Department of Physics, University of Oxford, Denys Wilkinson Building, Keble Road, Oxford, OX1 3RH, UK}
\email{jayne.birkby@physics.ox.ac.uk}

\author[orcid=0000-0003-4614-7035,sname='Biller']{Beth A. Biller}
\affiliation{Institute for Astronomy, University of Edinburgh, Royal Observatory, Blackford Hill, Edinburgh, EH9 3HJ, UK}
\affiliation{Centre for Exoplanet Science, University of Edinburgh, Edinburgh, EH9 3HJ, UK}
\email{bbiller@ed.ac.uk}

\author[orcid=0000-0001-9823-1445,sname='Dupuy']{Trent J. Dupuy}
\affiliation{Institute for Astronomy, University of Edinburgh, Royal Observatory, Blackford Hill, Edinburgh, EH9 3HJ, UK}
\affiliation{Centre for Exoplanet Science, University of Edinburgh, Edinburgh, EH9 3HJ, UK}
\email{tdupuy@roe.ac.uk}

\author[orcid=0000-0003-2530-9330,sname='Garvin']{Emily O. Garvin}
\affiliation{ETH Z\"urich, Institute for Particle Physics and Astrophysics, Wolfgang-Pauli-Strasse 27, 8093 Z\"urich, Switzerland}
\email{egarvin@phys.ethz.ch}

\author[orcid=0009-0000-8422-2222,sname='Iskandarli']{Leyla Iskandarli}
\affiliation{Breakthrough Listen, Department of Physics, University of Oxford, Oxford OX1 3RH, UK}
\email{iskandarlileyla@gmail.com}

\author[orcid=0000-0003-2769-0438,sname='Kammerer']{Jens Kammerer}
\affiliation{European Southern Observatory, Karl-Schwarzschild-Stra\ss{}e 2, 85748 Garching bei M\"unchen, Germany}
\email{jkammere@eso.org}

\author[orcid=0000-0003-2969-6040,sname='Zhou']{Yifan Zhou}
\affiliation{Department of Astronomy, University of Virginia, 530 McCormick Rd., Charlottesville, Virginia, 22904 USA}
\email{yzhou@virginia.edu}

\author[orcid=0000-0002-4918-0247,sname='De Rosa']{Robert J. De Rosa}
\affiliation{European Southern Observatory, Alonso de C\'{o}rdova 3107, Vitacura, Casilla 19001, Santiago, Chile}
\email{rderosa@eso.org}

\author[orcid=0000-0001-5365-4815,sname='Carter']{Aarynn L. Carter}
\affiliation{Space Telescope Science Institute, 3700 San Martin Dr, Baltimore, MD 21218, USA}
\email{aacarter@stsci.edu}

\author[orcid=0000-0001-8074-2562,sname='Hinkley']{Sasha Hinkley}
\affiliation{University of Exeter, Physics Building, Stocker Road, Exeter EX4 4QL, UK}
\email{S.Hinkley@exeter.ac.uk}

\author[orcid=0000-0002-7064-8270,sname='Kenworthy']{Matthew A. Kenworthy}
\affiliation{Leiden Observatory, Leiden University, Postbus 9513, 2300 RA Leiden, The Netherlands}
\email{kenworthy@strw.leidenuniv.nl}

\author[orcid=0000-0001-6396-8439,sname='Balmer']{William O. Balmer}
\affiliation{Department of Physics \& Astronomy, Johns Hopkins University, 3400 N. Charles Street, Baltimore, MD 21218, USA}
\email{wbalmer1@jh.edu}

\author[orcid=0000-0003-1502-4315,sname='Hammond']{Iain Hammond}
\affiliation{Max-Planck-Institut f\"ur Astronomie, K\"onigstuhl 17, D-69117 Heidelberg, Germany}
\email{iahammond@mpia.de}

\author[orcid=0000-0001-5864-9599,sname='Mang']{James Mang}
\altaffiliation{NSF Graduate Research Fellow.} 
\affiliation{Department of Astronomy, University of Texas at Austin, 2515 Speedway, Austin, TX 78712, USA}
\email{j_mang@utexas.edu}

\author[orcid=0000-0002-4404-0456,sname='Morley']{Caroline V. Morley}
\affiliation{Department of Astronomy, University of Texas at Austin, 2515 Speedway, Austin, TX 78712, USA}
\email{cmorley@utexas.edu}

\author[orcid=0009-0002-2455-7500,sname='Neeser']{Mark J. Neeser}
\affiliation{European Southern Observatory, Karl-Schwarzschild-Stra\ss{}e 2, 85748 Garching bei M\"unchen, Germany}
\email{mneeser@eso.org}

\author[orcid=0000-0002-4006-6237,sname='Absil']{Olivier Absil}
\altaffiliation{O.A. is a Research Director of the Fonds de la Recherche Scientifique -- FNRS}
\affiliation{STAR Institute, Universit\'e de Li\`ege, All\'ee du Six Ao\^ut 19c, 4000 Li\`ege, Belgium}
\email{olivier.absil@uliege.be}

\author[orcid=0000-0001-9353-2724,sname='Boccaletti']{Anthony Boccaletti}
\affiliation{LIRA, Observatoire de Paris, Universit\'e PSL, CNRS, Sorbonne Universit\'e, Univ. Paris Diderot, Sorbonne Paris Cit\'e, CY Cergy Paris Universit\'e, 5 place Jules Janssen, 92195 Meudon, France}
\email{anthony.boccaletti@obspm.fr}

\author[orcid=0000-0002-7520-8389,sname='Bonavita']{Mariangela Bonavita}
\affiliation{Institute for Astronomy, University of Edinburgh, Royal Observatory, Blackford Hill, Edinburgh, EH9 3HJ, UK}
\email{mbonavit@ed.ac.uk}

\author[orcid=0000-0003-2649-2288,sname='Bowler']{Brendan P. Bowler}
\affiliation{Department of Physics, University of California, Santa Barbara, Santa Barbara, CA 93106, USA}
\email{bpbowler@ucsb.edu}

\author[orcid=0009-0005-9339-2369,sname='Chen']{Xueqing Chen}
\affiliation{Institute for Astronomy, University of Edinburgh, Royal Observatory, Blackford Hill, Edinburgh, EH9 3HJ, UK}
\affiliation{Centre for Exoplanet Science, University of Edinburgh, Edinburgh, EH9 3HJ, UK}
\email{xueqing.chen@ed.ac.uk}

\author[orcid=0000-0002-5476-2663,sname='Dannert']{Felix A. Dannert}
\affiliation{ETH Z\"urich, Institute for Particle Physics and Astrophysics, Wolfgang-Pauli-Strasse 27, 8093 Z\"urich, Switzerland}
\email{fdannert@ethz.ch}

\author[orcid=0000-0001-8627-0404,sname='Girard']{Julien H. Girard}
\affiliation{Space Telescope Science Institute, 3700 San Martin Dr, Baltimore, MD 21218, USA}
\email{jgirard@stsci.edu}

\author[0000-0002-8425-6606,sname='Kasper']{Markus Kasper}
\affiliation{European Southern Observatory, Karl-Schwarzschild-Stra\ss{}e 2, 85748 Garching bei M\"unchen, Germany}
\email{kasper@mpia-hd.mpg.de}

\author[orcid=0000-0002-2189-2365,sname='Lagrange']{Anne-Marie Lagrange}
\affiliation{LIRA, Observatoire de Paris, Universit\'e PSL, CNRS, Sorbonne Universit\'e, Univ. Paris Diderot, Sorbonne Paris Cit\'e, CY Cergy Paris Universit\'e, 5 place Jules Janssen, 92195 Meudon, France}
\affiliation{Univ. Grenoble Alpes, CNRS, IPAG, F-38000 Grenoble, France}
\email{Anne-marie.Lagrange@obspm.fr}

\author[orcid=0000-0001-7047-0874,sname='Liu']{Pengyu Liu}
\affiliation{Institute for Astronomy, University of Edinburgh, Royal Observatory, Blackford Hill, Edinburgh, EH9 3HJ, UK}
\affiliation{Centre for Exoplanet Science, University of Edinburgh, Edinburgh, EH9 3HJ, UK}
\email{pengyu.liu@ed.ac.uk}

\author[orcid=0000-0002-4790-415X,sname='Orban de Xivry']{Gilles Orban de Xivry}
\affiliation{STAR Institute, Universit\'e de Li\`ege, All\'ee du Six Ao\^ut 19c, 4000 Li\`ege, Belgium}
\email{gorban@uliege.be}

\author[orcid=0000-0001-7739-9767,sname='Poon']{Michael Poon}
\affiliation{Department of Astronomy and Astrophysics, University of Toronto, 50 St. George Street, Toronto, ON M5S 3H4, Canada}
\email{michaelkm.poon@mail.utoronto.ca}

\author[orcid=0000-0003-3829-7412,sname='Quanz']{Sascha P. Quanz}
\affiliation{ETH Z\"urich, Institute for Particle Physics and Astrophysics, Wolfgang-Pauli-Strasse 27, 8093 Z\"urich, Switzerland}
\affiliation{ETH Z\"urich, Department of Earth and Planetary Sciences, Sonneggstrasse 5, 8092 Z\"urich, Switzerland}
\email{sascha.quanz@astro.phys.ethz.ch}

\author[orcid=0009-0005-5104-5772,sname='Serra']{Beno\^it Serra}
\affiliation{European Southern Observatory, Karl-Schwarzschild-Stra\ss{}e 2, 85748 Garching bei M\"unchen, Germany}
\email{benoit.serra@eso.org}

\author[orcid=0000-0003-0489-1528,sname='Vos']{Johanna M. Vos}
\affiliation{School of Physics, Trinity College Dublin, Dublin 2, Ireland}
\email{johanna.vos@tcd.ie}

\author[orcid=0000-0002-4309-6343,sname='Wagner']{Kevin Wagner}
\affiliation{Department of Astronomy and Steward Observatory, University of Arizona, 933 N. Cherry Ave., Tucson, AZ 85721, USA}
\email{kevinwagner@arizona.edu}

\author[orcid=0000-0003-0774-6502,sname='Wang']{Jason Wang}
\affiliation{Department of Physics and Astronomy, Northwestern University, 2145 Sheridan Road, Evanston, IL 60208-3112, USA}
\affiliation{Center for Interdisciplinary Exploration and Research in Astrophysics, 1800 Sherman Ave, Northwestern University, Evanston, IL 60201, USA}
\email{jason.wang@northwestern.edu}

\author[orcid=0000-0002-8177-0925,sname='Sch\"olkopf']{Bernhard Sch\"olkopf}
\affiliation{Max Planck Institute for Intelligent Systems, Max-Planck-Ring 4, 72076 T\"ubingen, Germany}
\affiliation{ELLIS Institute, Maria-von-Linden-Str. 2, 72076 T\"ubingen, Germany}
\email{bs@tuebingen.mpg.de}

\author[orcid=0000-0002-6382-2613,sname='Agapito']{Guido Agapito}
\affiliation{INAF -- Osservatorio Astrofisico di Arcetri, Largo E. Fermi 5., 50125, Firenze, Italy}
\email{guido.agapito@inaf.it}

\author[orcid=,sname='Agudo Berbel']{Alex Agudo Berbel}
\affiliation{Max Planck Institute for Extraterrestrial Physics, Giessenbachstra\ss{}e 1, 85748 Garching, Germany}
\email{agudo@mpe.mpg.de}

\author[orcid=0000-0003-3714-5855,sname='Apai']{D\'aniel Apai}
\affiliation{Department of Astronomy and Steward Observatory, University of Arizona, 933 N. Cherry Ave., Tucson, AZ 85721, USA}
\affiliation{Lunar and Planetary Laboratory, The University of Arizona, Tucson, AZ 85721, USA}
\affiliation{Department of Earth, Atmospheric and Planetary Sciences, Massachusetts Institute of Technology, Cambridge, MA 02139, USA}
\email{apai@arizona.edu}

\author[orcid=0000-0002-1114-4355,sname='Baruffolo']{Andrea Baruffolo}
\affiliation{INAF -- Osservatorio Astronomico di Padova, Vicolo dell’Osservatorio 5, 35122, Padova, Italy}
\email{andrea.baruffolo@inaf.it}

\author[orcid=,sname='Black']{Martin Black}
\affiliation{STFC UK Astronomy Technology Centre, Royal Observatory, Blackford Hill, Edinburgh, EH9 3HJ, UK}
\email{martin.black@stfc.ac.uk}

\author[orcid=0000-0002-4236-2339,sname='Bonaglia']{Marco Bonaglia}
\affiliation{INAF -- Osservatorio Astrofisico di Arcetri, Largo E. Fermi 5., 50125, Firenze, Italy}
\email{marco.bonaglia@inaf.it}

\author[orcid=0000-0002-0495-0543,sname='Briguglio']{Runa Briguglio}
\affiliation{INAF -- Osservatorio Astrofisico di Arcetri, Largo E. Fermi 5., 50125, Firenze, Italy}
\email{runa.briguglio@inaf.it}

\author[orcid=0000-0001-5301-1326,sname='Cao']{Yixian Cao}
\affiliation{Max Planck Institute for Extraterrestrial Physics, Giessenbachstra\ss{}e 1, 85748 Garching, Germany}
\email{ycao@mpe.mpg.de}

\author[orcid=0000-0003-1492-1591,sname='Carbonaro']{Luca Carbonaro}
\affiliation{INAF -- Osservatorio Astrofisico di Arcetri, Largo E. Fermi 5., 50125, Firenze, Italy}
\email{luca.carbonaro@inaf.it}

\author[orcid=,sname='Chapman']{Lee Chapman}
\affiliation{STFC UK Astronomy Technology Centre, Royal Observatory, Blackford Hill, Edinburgh, EH9 3HJ, UK}
\email{lee.chapman@stfc.ac.uk}

\author[orcid=0000-0002-5281-1417,sname='Cresci']{Giovanni Cresci}
\affiliation{INAF -- Osservatorio Astrofisico di Arcetri, Largo E. Fermi 5., 50125, Firenze, Italy}
\email{giovanni.cresci@inaf.it}

\author[orcid=,sname='Dallilar']{Yigit Dallilar}
\affiliation{I. Physikalisches Institut, Universit\"at zu K\"oln, Z\"ulpicher Stra\ss{}e 77, 50937, K\"oln, Germany}
\email{dallilar@ph1.uni-koeln.de}

\author[orcid=0000-0003-4949-7217,sname='Davies']{Richard Davies}
\affiliation{Max Planck Institute for Extraterrestrial Physics, Giessenbachstra\ss{}e 1, 85748 Garching, Germany}
\email{davies@mpe.mpg.de}

\author[orcid=,sname='Deysenroth']{Matthias Deysenroth}
\affiliation{Max Planck Institute for Extraterrestrial Physics, Giessenbachstra\ss{}e 1, 85748 Garching, Germany}
\email{m.deysenroth@mpe.mpg.de}

\author[orcid=0000-0003-3260-4389,sname='Di Antonio']{Ivan Di Antonio}
\affiliation{INAF -- Osservatorio Astronomico d’Abruzzo, Via Mentore Maggini, 64100, Teramo, Italy}
\email{ivan.diantonio@inaf.it}

\author[orcid=0000-0003-4508-5645,sname='Di Cianno']{Amico Di Cianno}
\affiliation{INAF -- Osservatorio Astronomico d’Abruzzo, Via Mentore Maggini, 64100, Teramo, Italy}
\email{amico.dicianno@inaf.it}

\author[orcid=0000-0002-5213-8269,sname='Di Rico']{Gianluca Di Rico}
\affiliation{INAF -- Osservatorio Astronomico d’Abruzzo, Via Mentore Maggini, 64100, Teramo, Italy}
\email{gianluca.dirico@inaf.it}

\author[orcid=0000-0003-0695-0480,sname='Doelman']{David Doelman}
\affiliation{Leiden Observatory, Leiden University, Postbus 9513, 2300 RA Leiden, The Netherlands}
\email{doelman@strw.leidenuniv.nl}

\author[orcid=0000-0001-8000-5642,sname='Dolci']{Mauro Dolci}
\affiliation{INAF -- Osservatorio Astronomico d’Abruzzo, Via Mentore Maggini, 64100, Teramo, Italy}
\email{mauro.dolci@inaf.it}

\author[orcid=,sname='Eisenhauer']{Frank Eisenhauer}
\affiliation{Max Planck Institute for Extraterrestrial Physics, Giessenbachstra\ss{}e 1, 85748 Garching, Germany}
\email{eisenhauer@mpe.mpg.de}

\author[orcid=0000-0002-3114-677X,sname='Esposito']{Simone Esposito}
\affiliation{INAF -- Osservatorio Astrofisico di Arcetri, Largo E. Fermi 5., 50125, Firenze, Italy}
\email{simone.esposito@inaf.it}

\author[orcid=0000-0003-1475-5976,sname='Ferruzzi']{Debora Ferruzzi}
\affiliation{INAF -- Osservatorio Astrofisico di Arcetri, Largo E. Fermi 5., 50125, Firenze, Italy}
\email{debora.ferruzzi@inaf.it}

\author[orcid=,sname='Feuchtgruber']{Helmut Feuchtgruber}
\affiliation{Max Planck Institute for Extraterrestrial Physics, Giessenbachstra\ss{}e 1, 85748 Garching, Germany}
\email{fgb@mpe.mpg.de}

\author[orcid=,sname='F\"orster-Schreiber']{Natascha F\"orster-Schreiber}
\affiliation{Max Planck Institute for Extraterrestrial Physics, Giessenbachstra\ss{}e 1, 85748 Garching, Germany}
\email{forster@mpe.mpg.de}

\author[orcid=0000-0003-4557-414X,sname='Franson']{Kyle Franson}
\altaffiliation{NASA Sagan Fellow}
\affiliation{Department of Astronomy \& Astrophysics, University of California, Santa Cruz, 1156 High Street, Santa Cruz, CA 95064, USA}
\email{kfranson@ucsc.edu}

\author[orcid=,sname='Genzel']{Reinhard Genzel}
\affiliation{Max Planck Institute for Extraterrestrial Physics, Giessenbachstra\ss{}e 1, 85748 Garching, Germany}
\email{genzel@mpe.mpg.de}

\author[orcid=,sname='Gillessen']{Stefan Gillessen}
\affiliation{Max Planck Institute for Extraterrestrial Physics, Giessenbachstra\ss{}e 1, 85748 Garching, Germany}
\email{ste@mpe.mpg.de}

\author[orcid=0000-0003-4636-6676,sname='Gonzales']{Eileen C. Gonzales}
\affiliation{Department of Physics and Astronomy, San Francisco State University, 1600 Holloway Ave., San Francisco, CA 94132, USA}
\email{egonzales@sfsu.edu}

\author[orcid=,sname='Hartl']{Michael Hartl}
\affiliation{Max Planck Institute for Extraterrestrial Physics, Giessenbachstra\ss{}e 1, 85748 Garching, Germany}
\email{hartl@mpe.mpg.de}

\author[orcid=0000-0003-3768-5712,sname='Hayoz']{Jean Hayoz}
\affiliation{ETH Z\"urich, Institute for Particle Physics and Astrophysics, Wolfgang-Pauli-Strasse 27, 8093 Z\"urich, Switzerland}
\email{jeanhayoz94@gmail.com}

\author[orcid=,sname='Huber']{Heinrich Huber}
\affiliation{Max Planck Institute for Extraterrestrial Physics, Giessenbachstra\ss{}e 1, 85748 Garching, Germany}
\email{HHUBER@MPE.MPG.DE}

\author[orcid=0000-0002-1368-841X,sname='Keller']{Christoph Keller}
\affiliation{National Solar Observatory, 3665 Discovery Drive, Boulder, CO 80303, USA}
\email{ckeller@nso.edu}

\author[orcid=,sname='Kravchenko']{Kateryna Kravchenko}
\affiliation{Max Planck Institute for Extraterrestrial Physics, Giessenbachstra\ss{}e 1, 85748 Garching, Germany}
\email{kkravchenko@mpe.mpg.de}

\author[orcid=0000-0002-0834-6140,sname='Leisenring']{Jarron Leisenring}
\affiliation{Department of Astronomy and Steward Observatory, University of Arizona, 933 N. Cherry Ave., Tucson, AZ 85721, USA}
\email{jarronl@arizona.edu}

\author[orcid=,sname='Lightfoot']{John Lightfoot}
\affiliation{STFC UK Astronomy Technology Centre, Royal Observatory, Blackford Hill, Edinburgh, EH9 3HJ, UK}
\email{johnlightfoot1957@gmail.com}

\author[orcid=,sname='Lunney']{David Lunney}
\affiliation{STFC UK Astronomy Technology Centre, Royal Observatory, Blackford Hill, Edinburgh, EH9 3HJ, UK}
\email{david.lunney@stfc.ac.uk}

\author[orcid=0000-0003-0291-9582,sname='Lutz']{Dieter Lutz}
\affiliation{Max Planck Institute for Extraterrestrial Physics, Giessenbachstra\ss{}e 1, 85748 Garching, Germany}
\email{lutz@mpe.mpg.de}

\author[orcid=0009-0000-5761-3332,sname='Macintosh']{Mike Macintosh}
\affiliation{STFC UK Astronomy Technology Centre, Royal Observatory, Blackford Hill, Edinburgh, EH9 3HJ, UK}
\email{mike.macintosh@stfc.ac.uk}

\author[orcid=0000-0002-4803-2381,sname='Mannucci']{Filippo Mannucci}
\affiliation{INAF -- Osservatorio Astrofisico di Arcetri, Largo E. Fermi 5., 50125, Firenze, Italy}
\email{filippo.mannucci@inaf.it}

\author[orcid=0000-0003-3050-8203,sname='Metchev']{Stanimir Metchev}
\affiliation{Department of Physics \& Astronomy, Western University, 1151 Richmond St, London, ON N6A 3K7, Canada}
\email{smetchev@uwo.ca}

\author[orcid=,sname='Ott']{Thomas Ott}
\affiliation{Max Planck Institute for Extraterrestrial Physics, Giessenbachstra\ss{}e 1, 85748 Garching, Germany}
\email{ott@mpe.mpg.de}

\author[orcid=,sname='Pearson']{David Pearson}
\affiliation{STFC UK Astronomy Technology Centre, Royal Observatory, Blackford Hill, Edinburgh, EH9 3HJ, UK}
\email{david.pearson@stfc.ac.uk}

\author[orcid=0000-0001-9553-8804,sname='Puglisi']{Alfio Puglisi}
\affiliation{INAF -- Osservatorio Astrofisico di Arcetri, Largo E. Fermi 5., 50125, Firenze, Italy}
\email{alfio.puglisi@inaf.it}

\author[orcid=,sname='Rabien']{Sebastian Rabien}
\affiliation{Max Planck Institute for Extraterrestrial Physics, Giessenbachstra\ss{}e 1, 85748 Garching, Germany}
\email{srabien@mpe.mpg.de}

\author[orcid=,sname='Rau']{Christian Rau}
\affiliation{Max Planck Institute for Extraterrestrial Physics, Giessenbachstra\ss{}e 1, 85748 Garching, Germany}
\email{rau@mpe.mpg.de}

\author[orcid=0000-0001-5460-2929,sname='Riccardi']{Armando Riccardi}
\affiliation{INAF -- Osservatorio Astrofisico di Arcetri, Largo E. Fermi 5., 50125, Firenze, Italy}
\email{armando.riccardi@inaf.it}

\author[orcid=0000-0002-7502-2701,sname='Salasnich']{Bernardo Salasnich}
\affiliation{INAF -- Osservatorio Astronomico di Padova, Vicolo dell’Osservatorio 5, 35122, Padova, Italy}
\email{bernardo.salasnich@inaf.it}

\author[orcid=0000-0002-2125-4670,sname='Shimizu']{Taro Shimizu}
\affiliation{Max Planck Institute for Extraterrestrial Physics, Giessenbachstra\ss{}e 1, 85748 Garching, Germany}
\email{shimizu@mpe.mpg.de}

\author[orcid=0000-0003-1946-7009,sname='Snik']{Frans Snik}
\affiliation{Leiden Observatory, Leiden University, Postbus 9513, 2300 RA Leiden, The Netherlands}
\email{snik@strw.leidenuniv.nl}

\author[orcid=,sname='Sturm']{Eckhard Sturm}
\affiliation{Max Planck Institute for Extraterrestrial Physics, Giessenbachstra\ss{}e 1, 85748 Garching, Germany}
\email{sturm@mpe.mpg.de}

\author[orcid=0000-0002-2011-4924,sname='Su\'arez']{Genaro Su\'arez}
\affiliation{Department of Astrophysics, American Museum of Natural History, Central Park West at 79th Street, New York, NY 10024, USA}
\email{gsuarez@amnh.org}

\author[orcid=,sname='Tacconi']{Linda Tacconi}
\affiliation{Max Planck Institute for Extraterrestrial Physics, Giessenbachstra\ss{}e 1, 85748 Garching, Germany}
\email{linda@mpe.mpg.de}

\author[orcid=0000-0003-2278-6932,sname='Tan']{Xianyu Tan}
\affiliation{Tsung-Dao Lee Institute and School of Physics and Astronomy, Shanghai Jiao Tong University, 1 Lisuo Road, Shanghai 200127, China}
\email{xianyut@sjtu.edu.cn}

\author[orcid=,sname='Taylor']{William Taylor}
\affiliation{STFC UK Astronomy Technology Centre, Royal Observatory, Blackford Hill, Edinburgh, EH9 3HJ, UK}
\email{william.taylor@stfc.ac.uk}

\author[orcid=0009-0009-3690-0821,sname='Waring']{Christopher Waring}
\affiliation{STFC UK Astronomy Technology Centre, Royal Observatory, Blackford Hill, Edinburgh, EH9 3HJ, UK}
\email{chris.waring@stfc.ac.uk}

\author[orcid=0000-0002-5565-084X,sname='Xompero']{Marco Xompero}
\affiliation{INAF -- Osservatorio Astrofisico di Arcetri, Largo E. Fermi 5., 50125, Firenze, Italy}
\email{marco.xompero@inaf.it}

\begin{abstract}
We report the direct imaging discovery of a third exoplanet in the $\beta$~Pictoris system. We detect $\beta$~Pictoris~d ($\beta$~Pic~d) in non-coronagraphic observations obtained with VLT/ERIS as well as multi-epoch archival datasets from JWST/NIRCam and VLT/SPHERE. Astrometric measurements over an 11-year baseline demonstrate that it is consistent with a gravitationally-bound source with orbital motion. Joint multi-planet orbit fits of all three planets in the system yield a semi-major axis of \sau_d\,au and inclination \inc_d\,deg for planet d. $\beta$~Pic~d has a larger orbital semi-major axis than the other known planets in the system, but is coplanar with the inner two planets, and its orbit is consistent with sculpting the inner edge of the debris disk. $\beta$~Pic~d has a contrast of $\Delta L^{\prime}=12.11\pm0.15$\,mag, with colors and luminosity that closely match those of 51~Eri~b, another exoplanet in the $\beta$~Pictoris moving group. Its VLT/ERIS and JWST/NIRCam colors are distinct from those of free-floating planetary-mass objects of a similar age and temperature. Its red $F410M-F444W$ color indicates strong CO$_2$ absorption in its atmosphere and suggests significant enhancement in metals compared to free-floating objects. From the ATMO hot-start evolutionary models, we estimate an effective temperature of \teff_d\,K and mass of \mass_d\,\Mjup, which also closely matches similar estimates for 51~Eri~b. $\beta$~Pic~d is among the lowest-mass exoplanets imaged from the ground. This discovery highlights the deep sensitivity achievable with ground-based imaging in the mid-infrared and the discovery potential of future high-contrast observations with the Extremely Large Telescope.
\end{abstract}


\section{Introduction}\label{sec:intro}
Since the detection of the first directly imaged exoplanets such as $\beta$~Pictoris~b \citep{2010Sci...329...57L} and HR~8799~bcd \citep{2008Sci...322.1348M}, near-infrared (near-IR) direct imaging has revealed a population of dozens of young $>$3\,\Mjup\ planets with effective temperatures of 800--2000\,K \citep[see~][for~a~recent~review]{2023ASPC..534..799C}. Many of these planets reside in multi-planet systems \citep[e.g.][]{2017A&A...597L...2M, 2018A&A...617A..44K, 2019NatAs...3..749H, 2020ApJ...898L..16B, 2023A&A...671L...5H, 2025ApJ...990L...8V, 2025A&A...704A.221V, 2025ApJ...990L...9C, 2026ApJ..1000L..38L}, including both of the two emblematic first directly imaged exoplanetary systems; HR~8799~e \citep{2010Natur.468.1080M} was announced soon after the other three \object{HR~8799} planets, and $\beta$~Pictoris~c was detected indirectly in 2019 \citep{2019NatAs...3.1135L}. Following this trend, we report the discovery of a third giant exoplanet in the \object{$\beta$~Pictoris} system. It seems likely that other directly imaged exoplanetary systems harbor additional lower mass planets that are not detectable with current instrumentation.

Recent improvements in sensitivity in mid-infrared instrumentation have enabled high-contrast imaging observations of a growing number of cold ($\lesssim$700\,K) giant exoplanets at 3--5\,$\mu$m wavelengths close to their peak blackbody emission, including the detections of \object{$\varepsilon$~Ind~A~b} \citep{2024Natur.633..789M}, \object{TWA~7~b} \citep{2025Natur.642..905L, 2025ApJ...987L..41C}, and \object{14~Her~c} \citep{2025ApJ...988L..18B} with JWST. At young ages ($<$100 Myr), this translates into the possibility to detect $<$3\,\Mjup\ exoplanets, which are more challenging to detect in the near-IR. While space-based facilities offer inherent stability and access to longer wavelengths, the largest ground-based telescopes remain highly complementary for mid-IR detections of such planets. This is in part because of the greater light-collecting power and angular resolution provided by their mirror sizes, which exceed that of JWST, e.g.\ the Very Large Telescope (VLT) Enhanced Resolution Imager and Spectrograph
\citep[ERIS;][]{2023A&A...674A.207D}.

With an age of \age_myr\,Myr \citep[][]{2024MNRAS.528.4760L} and distance of 19.63$\pm$0.06~pc \citep[][]{2016A&A...595A...1G, 2023A&A...674A...1G}, the A6V star \object{$\beta$~Pictoris} hosts two known exoplanetary companions, \object{$\beta$~Pictoris~b} \citep[\object{$\beta$~Pic~b};][]{2010Sci...329...57L} and \object{$\beta$~Pictoris~c} \citep[\object{$\beta$~Pic~c};][]{2019NatAs...3.1135L, 2020A&A...642L...2N}. The outer planet, $\beta$~Pic~b, has a dynamical mass of 9.3\raisebox{0.5ex}{\tiny$\substack{+2.6 \\ -2.5}$}\,\Mjup\ and a semi-major axis of 10.26$\pm$0.10\,au \citep{2020A&A...642A..18L, 2021AJ....161..179B}, while $\beta$~Pic~c has a comparable dynamical mass of 8.89$\pm$0.75\,\Mjup\ \citep{2025A&A...704A.318K} but a much closer-in orbital semi-major axis of 2.68$\pm$0.02\,au \citep{2021A&A...654L...2L}. Photometric and multi-resolution spectroscopic observations of $\beta$~Pic~b find a partially cloudy atmosphere with water and carbon monoxide, $\Teff\sim1700$\,K, $\logg\sim4.2$\,dex, and rotational velocity $v\sin(i)\sim20$ km~s$^{-1}$ \citep[e.g.,][]{2013A&A...555A.107B, 2013ApJ...776...15C, 2015ApJ...815..108M, 2017AJ....153..182C, 2020A&A...633A.110G, 2020A&A...635A.182S, 2024AJ....168...51K, 2024MNRAS.531.2356P, 2024A&A...682A..48L, 2025A&A...704A.182H, 2025A&A...704A.325R}. Spectroscopy in the $K$-band for $\beta$~Pic~c obtained with VLTI/GRAVITY \citep{2025A&A...704A.318K} suggests a similar effective temperature and early-L spectral type as $\beta$~Pic~b.

\begin{deluxetable*}{l l c c r c c c c c c c}
\tabletypesize{\scriptsize}
\tablecaption{List of observations of $\beta$~Pic used in this work, as well as the astrometry and photometry of $\beta$~Pic~d detections. \label{tab:obs}}
\tablecolumns{12}
\tablehead{
\colhead{ID}           & \colhead{Instrument}  & \colhead{Obs. Date} & \colhead{Filter}  & \colhead{Exp.} & \colhead{Seeing}   & \colhead{Field rot.}     & \colhead{S/N}     & \colhead{Separation}     & \colhead{Position Angle}   & \colhead{$\Delta$Mag.}  & \colhead{App. Mag.} \\
\colhead{}             & \colhead{}            & \colhead{(YMD-UT)}     & \colhead{}        & \colhead{(s)}      & \colhead{(\arcsec)}     & \colhead{(deg)}          & \colhead{}        & \colhead{(mas)}          & \colhead{(deg)}            & \colhead{(mag)}         & \colhead{(mag)}
}
\startdata
\textbf{1}   & VLT/ERIS      & 2025-12-03 & \Lp   & 7\,395   & 0.71    & 63.9  & 14.9 & 1126.9$\pm$3.9   & 210.35$\pm$0.22 & 12.11$\pm$0.15 & 15.57$\pm$0.16 \\
\textbf{2}   & JWST/NIRCam   & 2025-03-22 & $F410M$  & 47\,040  & \nodata    & 9.9   & 5.1  & 1104$\pm$20   & 211.1$\pm$1.0   & 11.6$\pm$0.2   & 15.2$\pm$0.2 \\
\textbf{3}   & JWST/NIRCam   & 2023-03-18 & $F444W$  & 3\,667   & \nodata    & 10.2  & 6.6  & 1012$\pm$20   & 210.4$\pm$1.0  & 11.6$\pm$0.2   & 14.7$\pm$0.2 \\
\textbf{4}   & VLT/SPHERE    & 2020-02-08 & $H3$    & 2\,024   & 0.60 & 30.7  & 5.9  & 784$\pm$15        & 209.4$\pm$1.0   & $<$15.7        & $<$19.2 \\
\textbf{5}   & VLT/SPHERE    & 2019-03-11 & $K2$    & 2\,808   & 0.72 & 32.1  & 6.2  & 719$\pm$12        & 209.6$\pm$1.0   & $<$14.9        & $<$18.4 \\
\textbf{6}    & VLT/SPHERE    & 2014-12-08 & $K1$    & 1\,980   & 0.77    & 26.1  & 7.9  & 375$\pm$20        & 208.0$\pm$2.0   & 14.1$\pm$0.7   & 17.6$\pm$0.7 \\
    &    &  & $K2$    &   &  &  & 7.2  &         &    & 13.6$\pm$0.6   & 17.1$\pm$0.6 \\
\hline 
\textbf{A}    & VLT/NaCo      & 2011-12-11 & \Lp   & 4800        & 0.57    & 54.1     & \nodata    & \nodata                 & \nodata               & \nodata              & \nodata \\
\enddata
\tablecomments{
 Epochs with a detection are assigned an ID number for convenient reference; the VLT/NaCo non-detection is indicated with `\textbf{A}'. For the JWST/NIRCam datasets, field rotation refers to the angular difference in roll positions. The S/N values for the ground-based observations are corrected for small number statistics \citep{Mawet2014,2023AJ....166...71B}, while we report \texttt{spaceKLIP} values for JWST/NIRCam \citep{2022SPIE12180E..3QG, 2022SPIE12180E..3NK, 2023ApJ...951L..20C}. Due to SDI self-subtraction, limits are reported for datasets \textbf{4} and \textbf{5}. Additional details are provided in the main text.
}
\end{deluxetable*}

$\beta$~Pictoris also possesses a large edge-on circumstellar debris disk \citep[spanning~between~$\sim$30-200~au~from~the star,][]{1984Sci...226.1421S, 2024AJ....167...69R,Han2026} that has long been connected to potential exoplanet formation in this system  \citep[e.g.][]{1993CeMDA..56..381S, 1994Icar..108...59L, 1995AAS...187.3205B, Mouillet1997, 2000ApJ...539..435H, 2003ApJ...584L..27W, Dent2014, 2023MNRAS.519.3257H, 2023A&A...675A..35S}. The known planets are believed to have played a significant role in sculpting disk structures, in particular, the significant warp observed in the $\beta$ Pictoris disk, with the inner \mbox{$\sim$80 au} tilted by $\sim5$\,deg relative to the outer disk \citep{1995AAS...187.3205B, Golimowski2006}. Both planets reside in near edge-on orbits coplanar with each other, although neither transit \citep{2019A&A...626A..97V, 2021A&A...648A..15K}. The orbits of these planets are thought to be inclined relative to the outer disk, and hence drive the warp \citep{Mouillet1997, Smallwood2023}. Recently, \citet{2025A&A...694A.236L} proposed the existence of an additional planet in the $\beta$~Pic system to explain the observed inner debris disk edge at 30--50\,au \citep{Han2026}. They found that the two known planets are unable to sculpt the disk out to this radius, and that a planet capable of doing so would necessarily reside beyond the orbit of $\beta$~Pic~b.

In this letter, we present the discovery of a third, wider-separation planet in the $\beta$~Pictoris system, \object{$\beta$~Pictoris~d} (hereinafter \object{$\beta$~Pic~d}), initially identified in non-coronagraphic direct imaging observations with VLT/ERIS. We confirm the planet through additional detections in earlier epochs of archival datasets from other instruments. We describe the corresponding data reduction methodology in Section~\ref{sect:obs_data_reduction}. In Section~\ref{sect:results}, we use astrometry to confirm that the observed source is gravitationally-bound to its host star and fit its orbital motion. We also use photometric measurements to begin characterising the atmosphere of the planet and estimate its mass. We then discuss the implications of the results for the $\beta$~Pic system in Section~\ref{sect:discussion}. Lastly, we summarize this work in Section~\ref{sect:conclusions}. An independent and contemporaneous discovery of $\beta$~Pictoris~d using JWST/NIRSpec+MIRI observations is separately reported by \citet{2026arXiv260623789G}.

\section{Observations and Data Reduction}\label{sect:obs_data_reduction}
We present results derived from observations obtained with four different instruments over 11~years. A summary of these observations can be found in Table~\ref{tab:obs}.
\begin{figure*}
    \centering
	\includegraphics[width=0.98\linewidth]{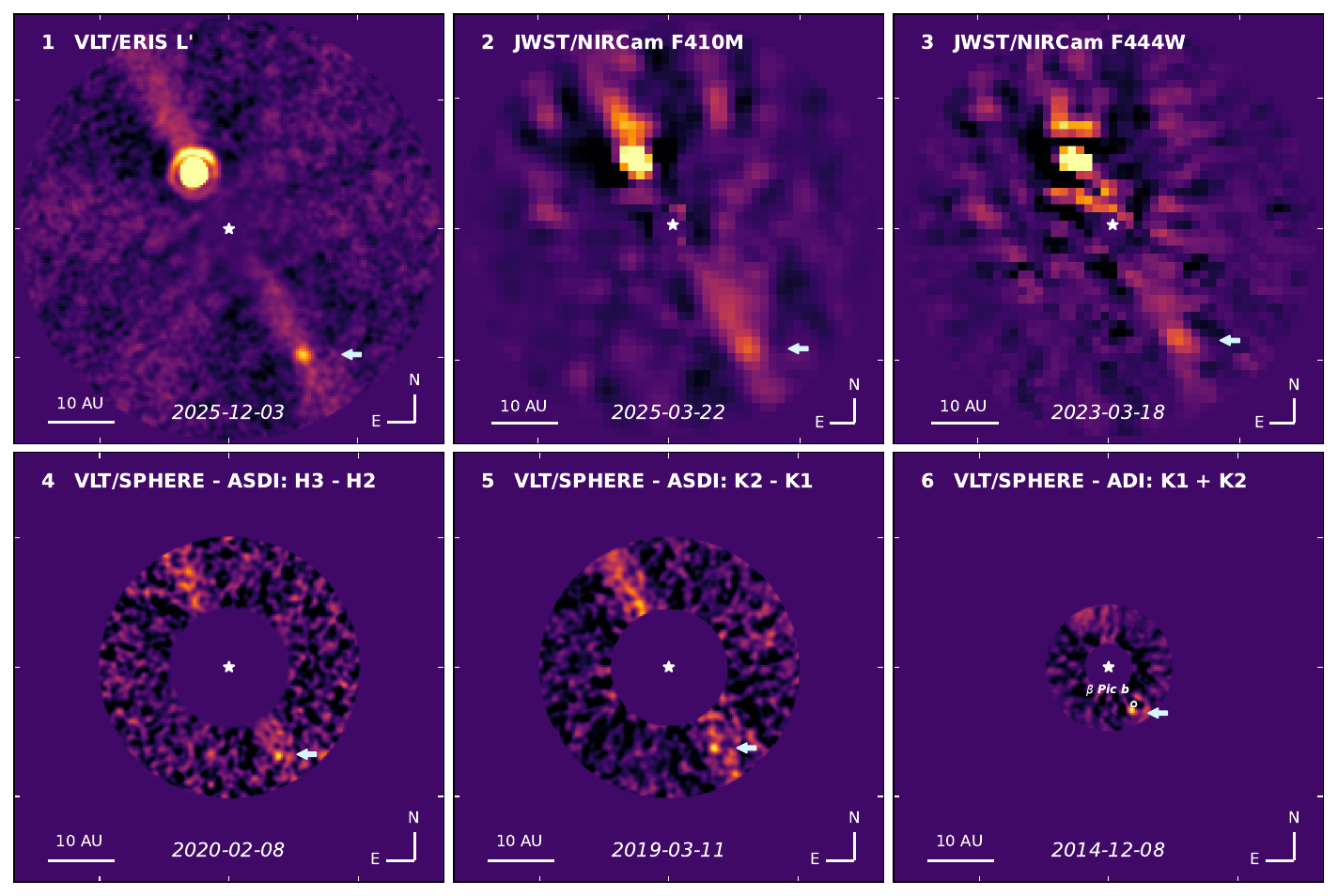}
    \caption{Gallery of images for the epochs at which $\beta$~Pic~d is detected (source indicated with a white arrow). The JWST/NIRCam images have been processed with a high-pass filter to reduce disk flux. The known planet $\beta$~Pic~b is the bright source in the North-East of the image in epochs 1--3. In epochs 4 and 5, $\beta$~Pic~b is within the inner mask used for data reduction. We note that for the 2014-12-08 epoch (epoch 6), $\beta$~Pic~b and $\beta$~Pic~d are nearly coincident; planet $\beta$~Pic~b (open circle) has been subtracted here to allow the much fainter $\beta$~Pic~d (arrow) to be seen. All images use the same normalized color scale and field of view, with ticks spaced at 1~arcsec intervals.}
    \label{fig:final_images}
\end{figure*}

\subsection{VLT/ERIS Observations}\label{obs_VLT_ERIS}
We observed the $\beta$~Pictoris system on 2025 December 03 (02:15:33--05:18:07~UT) with the ERIS Near Infrared Camera System \citep[NIX;][]{2016SPIE.9908E..3FP} on the VLT as part of a UK Guaranteed Time Observations programme with ESO programme no.\ 116.28WQ.001 (PI: Birkby, Biller). We used the 13\,mas\,pixel$^{-1}$ camera and the \Lp-band filter, which has a central wavelength of 3.79~$\mu$m, bandwidth of 0.60~$\mu$m, and zeropoint flux density of 239.24~Jy in the Vega system\footnote{Full filter information at: \url{https://svo2.cab.inta-csic.es/theory/fps/index.php?id=Paranal/ERIS.Lp}}. 
To maximize sensitivity, we used the LM-pupil without a coronagraph. Compared to the Lyot stop of the vortex coronagraph \citep{2024SPIE13097E..15O}, this provides better core throughput, enabling deeper background limits. 
The observations were obtained in pupil-stabilised mode, which permits the use of Angular Differential Imaging \citep[ADI;][]{2006ApJ...641..556M} for point spread function (PSF) subtraction.
We acquired 212 science frames on-target, each consisting of 4000 coadded frames with an integration time of 8.7214~ms per frame, for a total of 7395~s ($\sim$2~h) after removing 4 frames with poor AO correction. 
We used nodding to facilitate background subtraction while avoiding regions of clustered bad pixels.
These short individual exposure times ensured that the images remained unsaturated, which is essential to minimize artifacts caused by detector cross-talk and potential detector persistence. 

We reduced the data using \texttt{eristools}, a \texttt{Python} package for processing \eris HCI data based on \texttt{PynPoint} \citep{2019A&A...621A..59S}. The pipeline will be described in detail in a forthcoming publication (Bonse~et~al.~in~prep.). Using \texttt{eristools}, we perform the following steps:
First, we subtract dark frames and apply a flat-field correction using sky flats obtained at similar airmasses to those of the science frames. VLT/ERIS data also require correction for detector non-linearity.
We use \texttt{esorex} \citep{2015ascl.soft04003E} to fit a second-order polynomial to each individual pixel. 
The resulting non-linearity coefficients are imported into \texttt{eristools} to correct for the detector non-linearity.
Second, we subtract the background from each science frame using the frames at the opposite nod position.
\texttt{eristools} identifies bad pixels in two steps. First, we generate a bad-pixel map from the detector non-linearity coefficients. We then identify any remaining outliers through a $4\sigma$-clipping within a local region of size $5\times5$ pixels. 
In the final step, we aligned all frames relative to each other using sub-pixel cross-correlations, then centred them by fitting the unsaturated PSF with a Gaussian. 

For the post-processing, we applied a full-frame Principal Component Analysis \citep[PCA;][]{2012MNRAS.427..948A,2012ApJ...755L..28S} using \texttt{PynPoint} and \fours \citep{2025AJ....169..194B}, c.f., \citet[][]{2022A&A...666A...9G}. 
We explored different sets of processing parameters.
For PCA we calculated residuals with 1 to 50 principal components.
For the \fours reductions, the images were spatially downsampled by a factor of $1.5$ to keep the computational requirements manageable, and we varied the regularization parameter $\lambda_{\text{reg}}$ between $10^{5}$ and $5 \times 10^{6}$. The final residual image shown in Figure~\ref{fig:final_images} shows the \fours reduction with $\lambda_{\text{reg}} = 10^{5}$.

\subsection{JWST/NIRCam Observations}\label{obs_JWST_NIRCam}

We use two archival epochs of coronagraphic imaging observations targeting $\beta$~Pictoris with the JWST \citep{2006SSRv..123..485G, 2023PASP..135f8001G} Near Infrared Camera \citep[NIRCam;][]{2003SPIE.4850..478R, 2023PASP..135b8001R}.

The first of these was obtained on 2023 March 18~UT as part of GTO~1411 and was previously described by \citet{2024AJ....168...51K}. This dataset consists of observations in six filters, two in NIRCam's short-wavelength channel ($F182M$, $F210M$) using the MASK210R coronagraphic occulting mask, and four in its long-wavelength channel ($F250M$, $F300M$, $F335M$, $F444W$) using the MASK335R mask. The short and long-wavelength observations have pixel scales of 31\,mas\,pixel$^{-1}$ and 63\,mas\,pixel$^{-1}$, respectively. The total integration in each filter is given in Table~\ref{tab:obs}, and in Appendix~\ref{sect:non_detections}. Each observation was obtained at two roll positions, separated by 10.2\,deg, to enable PSF subtraction through ADI, and observations of a guide star, \object{$\alpha$~Pictoris}, were also acquired to allow Reference-star Differential Imaging \citep[RDI; e.g.][]{1998PASP..110.1046K}. The data reduction process for this dataset was carried out using the \texttt{pyKLIP} \citep{2015ascl.soft06001W} and \texttt{spaceKLIP} \citep{2022SPIE12180E..3QG, 2022SPIE12180E..3NK, 2023ApJ...951L..20C} pipelines and is described in detail in \citet{2024AJ....168...51K}.

The second JWST/NIRCam epoch was obtained on 2025 March 21--22 (11:23:20--03:34:07~UT) as part of GO~4758 (Zhou et al.\ 2026, accepted). These observations were acquired using the $F210M$ and $F410M$ filters in dual-band coronagraphic imaging mode and using the MASK335R occulting mask. Each exposure covered 2,352\,s of integration time. This was repeated 10 times for each of two rolls, separated by 9.9\,deg, giving 23,520\,s per roll and 47,040\,s ($\sim$13.07\,h) in total. As with the previous epoch, PSF subtraction was performed using combined ADI and RDI using reference star observations of $\alpha$~Pictoris. The data reduction process for this dataset is described in detail in Zhou et al.\ (2026, accepted); however, this process followed the same steps as for the previous epoch \citet{2024AJ....168...51K}. In brief, \texttt{spaceKLIP} was used to apply the stage 1 and 2 JWST pipelines\footnote{\url{https://spaceklip.readthedocs.io/en/latest/tutorials/tutorial_NIRCam_reductions.html}} to produce flux-calibrated images, correct bad pixels, and recenter the images. For both epochs, a 5-pixel-wide high-pass filter was applied to the images prior to PSF subtraction with Karhunen-Loève Image Projection \citep[KLIP;][]{2012MNRAS.427..948A, 2012ApJ...755L..28S, 2016ApJ...824..117P} to remove smooth, spatially extended light from the debris disk. We performed the PSF subtraction for both epochs by using the \texttt{pyklippipeline.run\_obs} function within \texttt{spaceKLIP} to apply KLIP with ADI+RDI, with 3 annuli, 1 subsection, and 50 KL modes.

\subsection{VLT/SPHERE Observations}\label{obs_SPHERE}
We use three archival epochs obtained with the Spectro-Polarimetric High contrast imager for Exoplanet REsearch \citep[SPHERE;][]{2019A&A...631A.155B} on the VLT between 2014~December~8~UT and 2020~February~8~UT as part of ESO programmes 60.A-9255(A), 60.A-9382(A) and 0104.C-0418(F), published in \citet{Lagrange2019} and \citet{
2020A&A...642A..18L}. These data were all obtained with SPHERE's Infrared Dual-band Imager and Spectrograph \citep[IRDIS;][]{2008SPIE.7014E..3LD} in either the $K$~band using the K12 filter pair, or the $H$~band using the H23 filter pair (see Table~\ref{tab:obs}). The imaging was obtained in IRDIFS mode, meaning that simultaneous SPHERE Integral Field Spectrograph observations were also acquired \cite[IFS;][]{Claudi2008}. Out of 18 SPHERE observations of the system between 2014 and 2020, we focused on the three IRDIS sequences that were the most favourable for the detection of the $\beta$~Pic~d candidate based on observing conditions, total integration time, and field rotation. We do not detect the candidate in initial reductions of the accompanying IFS data and in additional IRDIS datasets (see details in Appendix~\ref{sect:non_detections}). Therefore, we leave a deeper investigation of these data for a subsequent paper. For the three imaging epochs, we consider the nominal pixel scale and true north values of 12.255\,mas\,pixel$^{-1}$ and $-1.75$\,deg, respectively, given their stability over time \citep{Maire2016, Maire2021} and their consistency with the values inferred with dedicated astrometric calibration observations presented in \citet{Lagrange2019, 2020A&A...642A..18L}. The satellite spots were present in the entirety of all considered observations, enabling $\sim$1.5mas centering precision for the star when processing the image sequence \citep{Maire2021}.

We reduced the SPHERE data using the \texttt{vcal-sphere} pipeline \citep{Christiaens2023b}, which leverages both the ESO recipe execution tool (v3.13.10) for basic calibration (dark subtraction, flat-fielding, linearity correction, bad pixel identification, cube building) and routines from the Vortex Image Processing for pre- and post-processing \citep[VIP;][]{GomezGonzalez2017, Christiaens2023}. The VIP routines involved in the pipeline carry out a PCA-based sky subtraction, bad pixel clump identification and correction, coronagraphic image realignment based on the satellite spots, bad frame trimming, distortion correction, modeling and subtraction of the stellar signals using PCA \citep{2012MNRAS.427..948A, 2012ApJ...755L..28S}.

For post-processing, we considered two strategies to disentangle planet and speckle signals: ADI alone, and spectral differential imaging \citep[SDI;][]{SparksFord2002} followed by ADI through a two-step approach akin to the SADI method presented in \citet[][]{Christiaens2019}. 
In both cases, we tested PCA models using between 1 and 100 principal components. For the final images shown in Figure~\ref{fig:final_images}, we median-combined the reductions obtained with selected ranges of principal components: 5--40 for the 2014 ADI-only epoch, 10--40 for the 2019 SDI+ADI epoch, and 20--60 for the 2020 SDI+ADI epoch. Taking this additional median combination helps to further remove residual speckle noise while preserving signal that is robust with varying numbers of principal components. We considered a single annular region for the 2019 and 2020 epochs, where the inner radius was set to $\sim$450~mas in order to mask planet~b, and the outer radius was limited to 1~arcsec so as to not significantly extend calculation time. For the 2019 and 2020 epochs, we used a threshold in rotation corresponding to a minimum motion of 1 FWHM in the middle of the annulus, while a smaller annulus (4-FWHM wide around the position of planet b) with no rotation threshold was considered for the 2014 epoch. For the latter, we obtained the final image after subtracting planet~b from the data cube and carrying out a number of tests to assess the impact of this subtraction on the parameters estimated for planet d (more details in Sec.~\ref{sect:detection_of_d} and Appendix~\ref{sect:SubtractionOfb}).

\subsection{VLT/NaCo Observations}\label{obs_NaCO}
We re-processed 10 archival \Lp-band datasets obtained with the NaCo \citep[][]{2003SPIE.4841..944L, 2003SPIE.4839..140R, 2012SPIE.8447E..0LG} instrument in the years between 2009 and 2015, as well as data from 2003 that led to the discovery of $\beta$~Pic~b \citep{2009A&A...493L..21L}. 
All datasets, except the early data from 2003, were taken with pupil tracking and allowed for ADI. No coronagraph was used for any observation.
For pre-processing we use an automated pipeline based on \texttt{PynPoint} \citep{2019A&A...621A..59S} to perform the following steps: First, standard calibrations (dark subtraction and flat-field correction) are applied, followed by the removal of the thermal background using averaged sky frames.
We correct bad pixels using a $4\sigma$-clipping algorithm within a $5\times5$ local region. 
Next, we align the frames via sub-pixel cross-correlation. Temporal binning is not performed prior to PSF subtraction in order to preserve temporal resolution and improve the alignment of frames. Finally, we use a PCA-based criterion to reject low-quality frames, including those affected by an open adaptive optics loop.
As with the ERIS data, we use PCA \citep{2012MNRAS.427..948A,2012ApJ...755L..28S} and \fours \citep{2025AJ....169..194B} for PSF-subtraction. 
We varied the number of principal components between 1 and 100, and tested regularization of \fours between $10^{4}$ and $10^{6}$.
For PSF subtraction of the 2003 data, we first subtract the reference star \object{HR~2435}, which was used in \citet{2009A&A...493L..21L}. We then create a pseudo-pupil-tracking dataset by rotating the frames based on their parallactic angle to align the spider pattern. Finally, we process the data in a similar manner to the newer NACO datasets using PCA.

\section{Results}\label{sect:results}

\subsection{Detections of \texorpdfstring{$\beta$~Pictoris~d}{beta Pictoris d}}\label{sect:detection_of_d}

We present the final images from the epochs at which we detect $\beta$~Pic~d in Figure~\ref{fig:final_images}. We detect a clear point source, the putative $\beta$~Pic~d, South-West of the star in the VLT/ERIS data with a S/N of 14.9. We detect a compatible point source in the longest wavelength filters for each of the two JWST/NIRCam epochs, and for the three VLT/SPHERE epochs, including detections in both the $K12$ and $H23$ bands. We do not detect the source in the VLT/NaCo epochs. We summarise all of the detections in Table~\ref{tab:obs}, and non-detections in Appendix~\ref{sect:non_detections}. 

As we progressively recovered the $\beta$~Pic~d candidate in more datasets, we used \texttt{Octofitter} \citep{Thompson_2023} and \texttt{orbitize!} \citep{2020AJ....159...89B, 2024JOSS....9.6756B} to fit preliminary single-planet orbits to the later epochs and predict the expected position of the candidate in earlier datasets. For the earliest SPHERE epoch, from 2014, these indicated that the $\beta$~Pic~d candidate would likely be very close to $\beta$~Pic~b. We therefore first subtracted the much brighter planet~b from the 2014 datacube using the position and flux found by the negative fake companion method \citep[NegFC;][]{2010Sci...329...57L, Wertz2017,2019A&A...621A..59S}. After removing $\beta$~Pic~b, we processed this cube with PCA-ADI and identified a point source located in the wings of the PSF of planet~b ($\sim$45\,mas apart). This detection at S/N$\sim7.5$ is significant \citep[][]{Mawet2014,2023AJ....166...71B}. It is also robust with respect to the uncertainties on the parameters of planet~b used for the subtraction, although the latter dominate the uncertainty budget for the parameters estimated for the planet d candidate. We provide additional details in Appendix~\ref{sect:SubtractionOfb}.

For the 2019 and 2020 VLT/SPHERE epochs, we combined SDI and ADI to achieve a deeper contrast. This was necessary as the $\beta$~Pic~d candidate falls close to the extreme-AO corrected radius of VLT/SPHERE. At these separations, fast varying speckles affect the contrast achieved with ADI only. We only detect the candidate when combining SDI and ADI to remove these speckles.

At the epochs of available VLT/NaCo observations (between 2003 and 2015), our preliminary orbit fits predicted the candidate to be $<$0.5\arcsec\ from the star. We therefore did not detect the candidate at these separations in any of the VLT/NaCo observations because it was too close to the star or much brighter planet~b. However, as the sensitivity of these images at wider angular separation was sufficient to recover a source with the \Lp-band flux of the $\beta$~Pic~d candidate, these observations remained highly valuable for ruling out the possibility that the point source was a background source (see Section~\ref{sect:results_astrometry}). 

\subsection{Astrometric and photometric measurements}\label{sect:astrom_photom}

We measured the astrometry and photometry of the source in the different datasets using NegFC as described in the following subsections and report these values in Table~\ref{tab:obs} for each detection.

\subsubsection{VLT/ERIS}
Due to the high computing requirements of \fours, we run the NegFC injection for ERIS based on PCA. 
A high-pass filter was applied similar to the JWST reduction to prevent contamination due to the disk signal.
The fitting error is very stable for 5--15 PCA components, with uncertainties of 0.6~mas in separation, 0.03~deg in position angle, and 0.02~mag in photometry.
In addition to the fitting errors from the NegFC injection, we consider the following sources of error for the astrometry and photometry: 1.\ a precision of $\sim$0.25 pixels on centering of the star; 2.\ an additional uncertainty in position angle of 0.023\,deg on the true north calibration (see ERIS User Manual\footnote{\url{https://www.eso.org/sci/facilities/paranal/instruments/eris/doc/ERIS_User_Manual_v117.1.pdf}}); 3.\ the variability of the unsaturated reference PSF, which contributes an uncertainty of 0.125\,mag.

\subsubsection{JWST/NIRCam} 
For the JWST/NIRCam measurements, we used the routines in \texttt{spaceKLIP}, following the method applied by \citet[][]{2023ApJ...951L..20C} and \citet{2024AJ....168...51K}. We used these routines to model the KLIP-processed PSF of the candidate accounting for companion self-subtraction, and fit these to the images using Markov Chain Monte Carlo (MCMC) sampling as integrated in \texttt{spaceKLIP} with \texttt{emcee} \citep{2013PASP..125..306F}. We adopt conservative errors on the astrometry to account for uncertainties in the position of the host star behind the coronagraphic mask and the distortion correction in spaceKLIP, as noted by \citet{2024AJ....168...51K}, which combined introduce systematic uncertainties of $\sim$10~mas. We combine this in quadrature with the statistical uncertainties from the MCMC fit and round up to the nearest 10~mas in $\Delta$RA and $\Delta$Dec. Although the disk flux was well removed by the high-pass filter during data reduction, we also adopt conservative errors on the photometry to account for the impact of any residual disk flux. We do this by measuring the approximate disk flux at the location of the $\beta$~Pic~d candidate and combining this as an error in quadrature with the uncertainties from the MCMC fit.

\subsubsection{VLT/SPHERE}
We detail our method for measuring the astrometry and photometry of the $\beta$~Pic~d candidate in the 2014 VLT/SPHERE data, for which $\beta$~Pic~b was first removed, in Appendix~\ref{sect:astrometry_photometry_2014}. We estimate contrasts of $\Delta K1 = 14.1 \pm 0.7$\,mag and $\Delta K2 = 13.6 \pm 0.6$\,mag at this epoch. The considered errors on both the astrometry and photometry are conservative due to the uncertainty caused by the proximity to $\beta$~Pic~b, which is $\sim$100 times brighter than $\beta$~Pic~d in $K1$ and $K2$.

For the 2019 and 2020 VLT/SPHERE epochs obtained by combining SDI and ADI, the uncertainty associated with the PA is dominated by residual speckle noise. We estimate this from the injection and recovery of 180 fake companions at the same flux and radial separation, but different azimuths, than that estimated for the planet d candidate. In comparison, the uncertainty associated with the true north is negligible \citep[0.04$\degr$;][]{Maire2021}. The uncertainty associated with the radial separation for these SDI+ADI detections is dominated by the poorly quantified amount of SDI self-subtraction. During SDI, we scale the shortward band image (both in physical size and in flux) to match the speckle pattern of the longward band. This involves a flux scaling and radial motion outward of the companion by $\sim$0.6-0.8 FWHM (at the 2019 and 2020 epochs), before subtraction. We adopted a conservative radial-separation uncertainty that scales inversely with the radial motion of the planet between the two channels. This term reflects the maximum centroid displacement that could be introduced by SDI self-subtraction, which we consider negligible once the radial motion is larger than 1~FWHM.
A dedicated 2-channel forward modeling approach is needed to accurately estimate the planet's astrometry and its photometry in both channels, which is beyond the scope of this Letter and will be presented in a subsequent paper. Here, we report a lower limit on the flux in the longest-wavelength band, either $K2$ or $H3$. This limit corresponds to the case where planet self-subtraction is negligible when the shorter-wavelength image is rescaled and subtracted from the longer-wavelength image before ADI.
Using NegFC provides a flux estimate that is corrected for ADI self-subtraction, hence leading to an overall lower (upper) limit in flux (magnitude) contrast in $K2$ and $H3$. For $K2$, this limit is consistent with the estimate from 2014 obtained with ADI alone.

\subsection{Common proper motion analysis}\label{sect:results_astrometry}
\begin{figure}
    \centering
	\includegraphics[width=\linewidth]{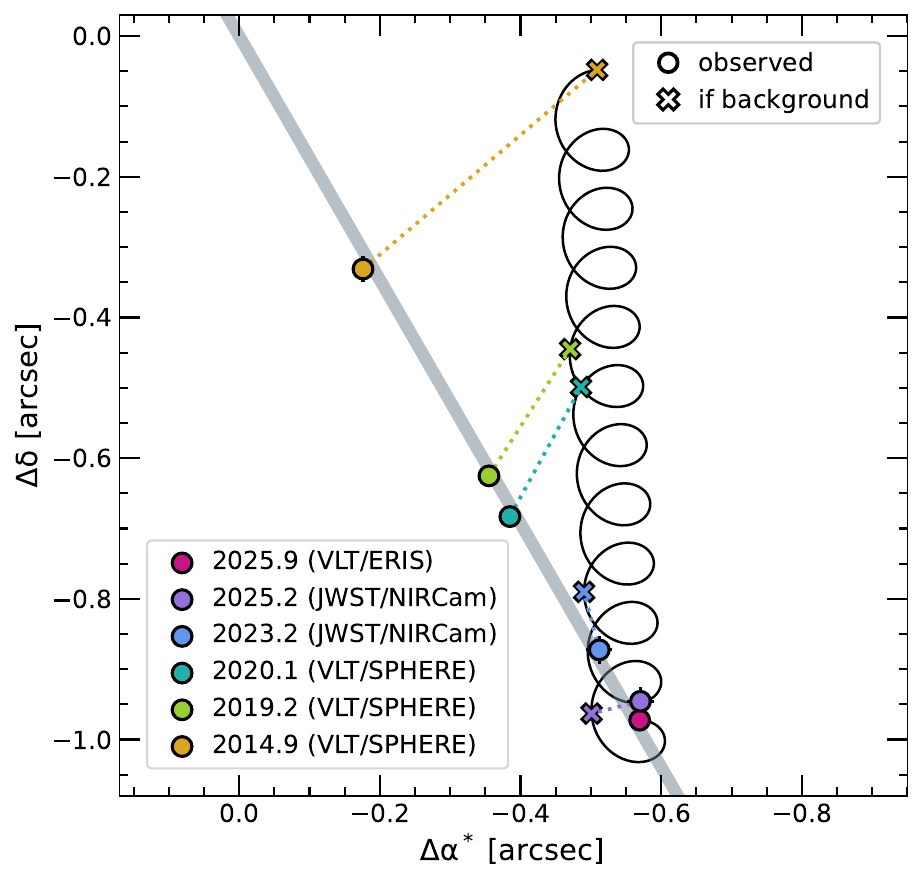}
    \caption{The astrometry of $\beta$~Pic~d relative to its host star at each epoch (circles) compared to its expected position if the candidate was a stationary background source (corresponding crosses). The spiral line indicates the expected motion of a stationary background source, as calculated using the Gaia~DR3 proper motion and parallax of $\beta$~Pic and projected backwards in time from the recent VLT/ERIS epoch (2025-12-03). The thick grey line represents the plane of the $\beta$~Pic debris disk \citep[PA$\sim$30~deg,][]{2024AJ....167...69R}.}
    \label{fig:cpm_detections}
\end{figure}

We plot the astrometry of the $\beta$~Pic~d candidate relative to its host star at each epoch in Figure~\ref{fig:cpm_detections}, alongside the corresponding positions that would be expected for a stationary background source at the same epochs, based on the parallax (50.93$\pm$0.15~mas) and proper motion ($\mu_\alpha=5.16\pm0.20\,\mathrm{mas}\,\mathrm{yr}^{-1}$, $\mu_\delta=84.04\pm0.19\,\mathrm{mas}\,\mathrm{yr}^{-1}$) of $\beta$~Pic \citep[Gaia DR3;][]{2016A&A...595A...1G, 2023A&A...674A...1G}. At all epochs, the measured position of the candidate differs significantly from that expected of a background object, with the overall motion across the $\sim$11-year baseline deviating from the background track at the $>$24$\sigma$ level. Furthermore, the motion of the candidate appears to be coplanar with the edge-on $\beta$~Pic debris disk (indicated with a thick grey line), as for the two known planets in the system.

As an additional check, we verified that no source was visible at the expected background location in the VLT/NaCo epochs taken in the \Lp\ filter.
Although we did not detect the $\beta$~Pic~d candidate in these datasets, the sensitivity of dataset (ID: \textbf{A}, Table~\ref{tab:obs}) is sufficient to detect a background source if one were present.
We inject an artificial source with $\Lp = 12.11$\,mag at the background position on 2011-12-11 and recover the signal at S/N of 6.3.
Therefore, we conclude that the candidate is not a background source and that the observed motion is that of a gravitationally bound companion moving outward in its orbit.

\subsection{Orbit fitting}\label{sect:orbit_fitting}

With an 11-year baseline and significant orbital motion observed across available epochs, we were able to perform an analysis of $\beta$~Pic~d's orbit and place initial constraints on its configuration. The detailed description of our orbit fitting procedure is presented in Appendix~\ref{A:orbit-fit}, and we briefly summarize the key points here. We performed a joint multi-planet analysis of $\beta$~Pic~b, c, and~d using \texttt{orvara} \citep[v1.1.2;][]{2021AJ....162..186B}, complementing our new relative astrometry for planet~d with literature measurements of the inner planets, along with radial velocity data and absolute astrometry of the star \footnote{We note that \texttt{orvara} reports the orbital parameters of the planets, not the star.}. Since planet~d is expected to have negligible influence on the star's acceleration, we conservatively chose to neglect the mass of $\beta$~Pic~d in the fits, thereby assuming that the RVs and absolute astrometry are not sensitive to planet~d (within measurement uncertainties), while still accounting for the motion of the star under the influence of the inner two planets when fitting the relative astrometry of planet~d. This was enforced by setting a Gaussian prior of $10^{-6}\pm10^{-9}$\,\Msun on planet~d's mass.

Results for the semi-major axis ($a$), eccentricity ($e$), and inclination ($i$) of planet~d are shown in Figure~\ref{fig:corner}, while the remaining orbital parameters are presented in Appendix~\ref{A:orbit-results}, along with results for planets~b and~c for which almost all orbital parameters are well constrained. The best constrained orbital elements for planet~d are the inclination and longitude of the ascending node, characterising the orientation of its orbit ($i_{\rm d}= \inc_d$\,deg, $\Omega_{\rm d}=\asc_d$\,deg), which indicate an orbit closely aligned with those of planets~b ($i_{\rm b}=88.993$\raisebox{0.5ex}{\tiny$\substack{+0.010 \\ -0.011}$}\,deg, $\Omega_{\rm b}=211.787$\raisebox{0.5ex}{\tiny$\substack{+0.010 \\ -0.008}$}\,deg) and~c ($i_{\rm c}=88.93$\raisebox{0.5ex}{\tiny$\substack{+0.11 \\ -0.09}$}\,deg, $\Omega_{\rm c}=211.08$\raisebox{0.5ex}{\tiny$\substack{+0.05 \\ -0.04}$}\,deg), despite adopting uninformed priors and without enforcing coplanarity with the known planetary system architecture. Since planet~d is constrained only by relative astrometry, our inferred value of $\Omega_{\rm d}$ is subject to the usual 180\,deg degeneracy of relative astrometric orbits, as the observations alone cannot distinguish whether the orbital motion is directed toward or away from the observer. We therefore restricted the fit to values of $\Omega_{\rm d}$ corresponding to the same direction of orbital motion as the other planets in the system. The alternative, degenerate solution corresponds to a coplanar but retrograde orbit and is obtained through the transformation $\lambda_{\rm d} \rightarrow \lambda_{\rm d} + 180$\,deg and $\omega_{\rm d} \rightarrow \omega_{\rm d} + 180$\,deg, with the corresponding shift $\Omega_{\rm d} \rightarrow \Omega_{\rm d} + 180$\,deg.

On the other hand, a strong covariance is seen between $e_{\rm d}$ and $a_{\rm d}$, with multiple families of solutions consistent with observations: low-$e$ orbits at moderate $a$; a branch of higher-$e$ orbits extending to lower $a$ and corresponding to our observation epochs being near apoastron; and a branch of higher-$e$ orbits at larger $a$ corresponding to our observation epochs being near periastron. As a result, although there is a weak preference for lower eccentricities, we can only place an upper limit on $e_{\rm d}<0.90$ at 2$\sigma$. Likewise, the semi-major axis posterior is bimodal, with peaks for the two sets of solutions at about 18\,au and 26\,au.

\begin{figure*}
    \centering
	\includegraphics[width=\linewidth]{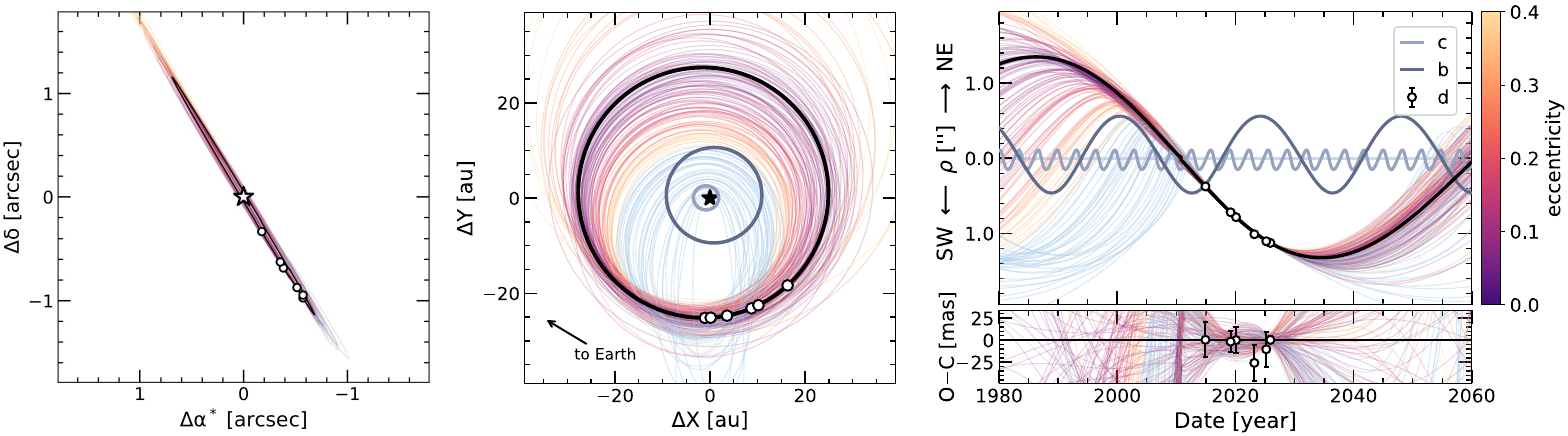}
    \caption{Left: astrometry of $\beta$~Pic~d (white circles) relative to the host star (white star). We show 150 randomly drawn samples from our posteriors, color-coded by eccentricity from low to high (purple to orange) unless the sample crosses the orbit of $\beta$~Pic~b (light blue). The solid black line indicates the lowest-$\chi^2$ orbit. Middle: the same information de-projected by the corresponding inclination for each orbit. The lowest-$\chi^2$ orbits for planets~b (dark grey) and~c (light grey) are plotted for comparison. Right: projected separation from the host star as a function of time for all three planets. The 2014 VLT/SPHERE point illustrates the near-coincident projected positions of planets~b and~d as seen from Earth.}
    \label{fig:orbit_fit}
\end{figure*}

In Figure~\ref{fig:orbit_fit} however, we compare our orbit fit to the relative astrometry measurements for planet~d, as well as a deprojected view of the $\beta$~Pic system. There is a subset of high-$e$, low-$a$ orbital solutions for planet~d that appear to cross the orbit of planet~b (and in some cases, even planet~c) and would thus likely be unstable. While the lack of mass measurement for planet~d prevents any advanced dynamical stability analysis at this stage, we used the analytical criterion of \citet[Eq.~6]{Whitmire1998}, which provides a geometric test for whether two coplanar orbits intersect for any eccentricities and orientations. From this, we determined that 28\% of the posterior solutions are in configurations where the orbits of planet~b and~d would intersect (light blue orbits in Figure~\ref{fig:orbit_fit}), making them likely unphysical. We thus discarded these solutions from our posterior samples. 

The final posteriors after removal of orbit-crossing solutions are shown in the darker contours in Figure~\ref{fig:corner}. Considering only the non-crossing orbits eliminates the highest-$e$ configurations of planet~d, resulting in $e_{\rm d}<0.44$ at 2$\sigma$. Without the highly-eccentric orbits, the semi-major axis distribution becomes unimodal, at \sau_d\,au. Table~\ref{tab:exoplanet_orbit_parameters} reports results for planet~d's main orbital elements derived from the final posterior distribution restricted to non-crossing solutions (see Appendix~\ref{A:orbit-results} for remaining orbital parameters). We note that $i_{\rm d}$ and $\Omega_{\rm d}$ are largely unaffected by the removal of orbit-crossing posterior solutions, as expected for truly independently constrained parameters. However, other orbital parameter posteriors are influenced by our Bayesian priors, particularly the uniform $e_{\rm d}$ prior, and therefore should be treated with caution.

\begin{figure}
    \centering
    \includegraphics[width=\linewidth]{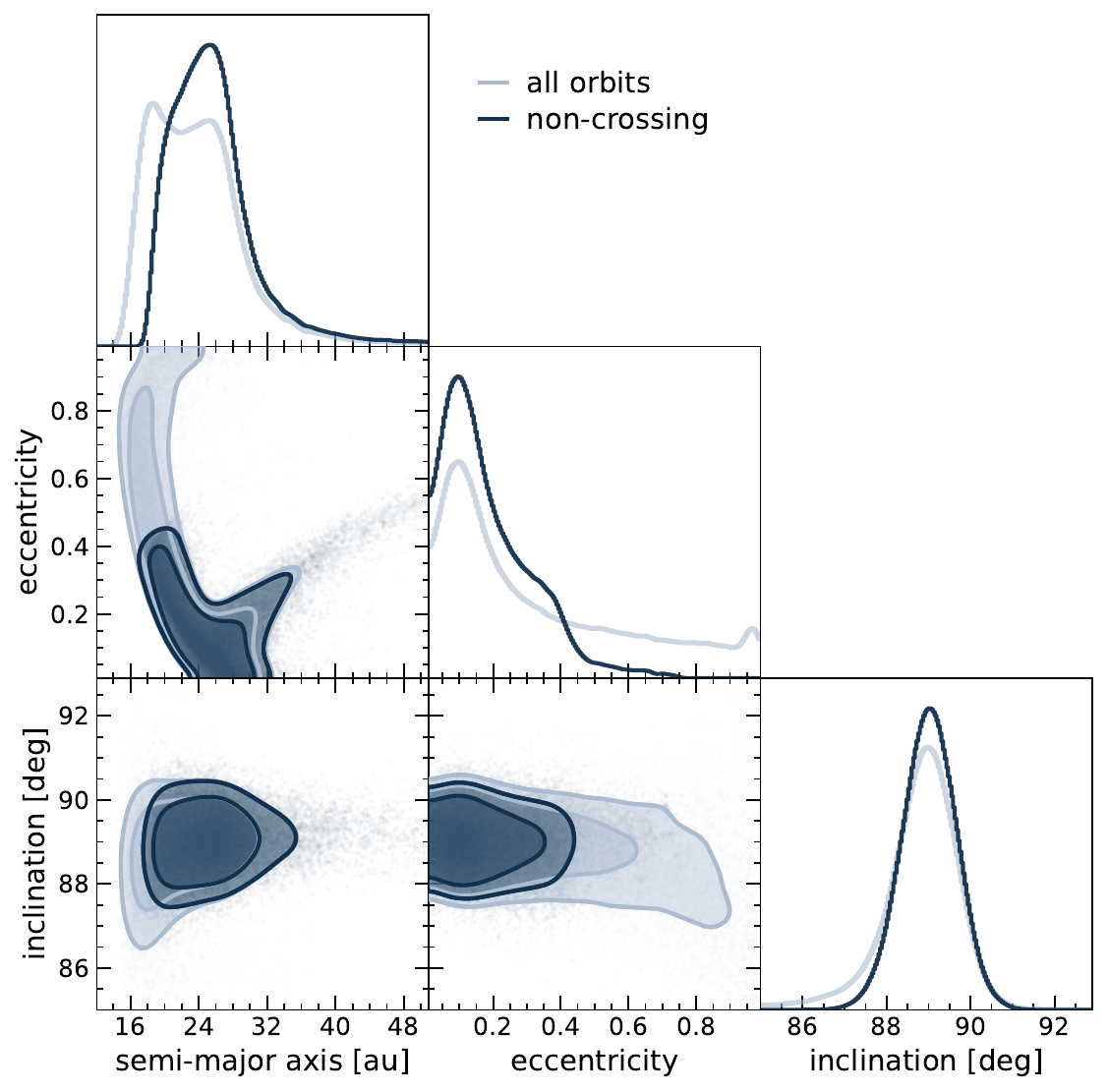}
    \hfill
    \caption{Marginalized posterior distributions for the semi-major axis, eccentricity, and inclination of planet~d from our orbital fit, along with their joint posteriors. Lighter and darker shades show the complete posterior samples and the subset with orbit-crossing solutions removed, respectively. Contours over the joint posteriors indicate the 1$\sigma$ and 2$\sigma$ credible regions.}
    \label{fig:corner}
\end{figure}

\begin{deluxetable}{lcc}
\tablecaption{Properties of $\beta$~Pic~d based on astrometric and photometric analysis.\label{tab:exoplanet_orbit_parameters}}
\tabletypesize{\normalsize}
\tablehead{
\colhead{Parameter} & \colhead{Value} & \colhead{Note}
}
\startdata
$a_{\rm d}$ (au)       &     \sau_d\phn     & non-crossing \\
$P_{\rm d}$ (yr)       &     \per_d         & non-crossing \\
$e_{\rm d}$            &     $<0.44$        & non-crossing, 2$\sigma$ c.i. \\
$i_{\rm d}$ (deg)      &     \inc_d\phn     & prior-independent \\
$\Omega_{\rm d}$ (deg) &     \asc_d\phn\phn & prior-independent \\
\hline
$m_{\rm d}$ (\Mjup)    &    ~\mass_d        & estimated \\
$T_{\rm eff, d}$ (K)   &     \teff_d\phn    & estimated \\
$R_{\rm d}$ (\Rjup)    &    ~\radius_d      & estimated \\
\enddata
\tablecomments{Orbital parameters are derived either from the posteriors with planet-b-orbit-crossing solutions removed or the normal analysis if the measurement is independent of priors.  Mass, \Teff, and radius estimates are from a combination of \texttt{ATMO}\,2020 chemical equilibrium and strong non-equilibrium evolutionary models, assuming an age of \age_myr\,Myr.}
\end{deluxetable}

\subsection{Photometric Analysis and Mass Estimate}\label{sect:colour}

\begin{figure*}
    \centering
	\includegraphics[width=\linewidth, trim=0.7cm 0.3cm 0.95cm 0.1cm, clip]{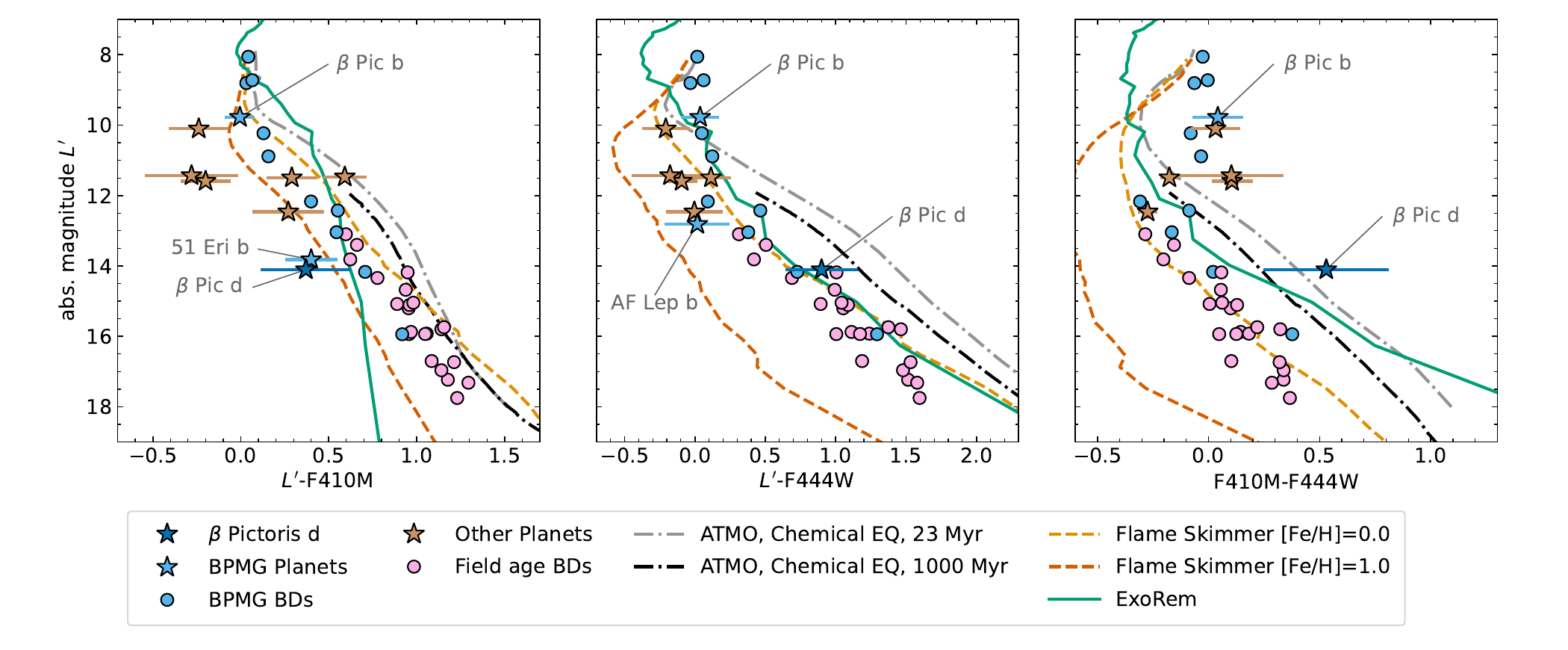}
    \caption{Color-magnitude diagrams of $\beta$~Pic~d, shown alongside photometry for directly imaged planets in BPMG (blue stars) and around other stars (brown stars), as well as free-floating brown dwarfs in BPMG (blue circles) and the field (pink circles). 
    While young (BPMG) and field-age brown dwarfs have similar colors as a function of magnitude, the planets are markedly fainter at F410M than brown dwarfs with comparable magnitude. We show evolutionary models from ATMO \citep{2020A&A...637A..38P} and atmosphere models from Sonora Flame Skimmer (Mang et al.~in prep.) and ExoRem \citep{2015A&A...582A..83B, 2018ApJ...854..172C}. For the atmosphere models, we assume $\log g=4.0$\,dex and $R=1.3\,\Rjup$. Flame Skimmer models adopt solar C/O, $K_{zz}=4$, and two metallicities ([Fe/H] = 0.0 and 1.0), while the ExoRem model has [Fe/H]=0.5, C/O=0.65, and is chosen based on the best-fit model of 51~Eri~b in \citet{2025AJ....169..209B}; this model also provides the closest match to $\beta$~Pic~d of all models shown.
    }
    \label{fig:CMD}
\end{figure*}

In this Section we use the color and luminosity of $\beta$~Pic~d to infer an effective temperature and mass, and study the planet's atmosphere. In Figure~\ref{fig:CMD}, we present color-magnitude diagrams comparing the ERIS and JWST photometry of $\beta$~Pic~d against other directly imaged planets, field brown dwarfs, and evolutionary model tracks, with young objects highlighted (see Appendix~\ref{app:CMD_literature} for details). There are five known planets and several free-floating brown dwarfs in the $\beta$~Pictoris moving group \citep[BPMG;][]{2001ApJ...562L..87Z}; all are co-eval and formed in a similar environment, providing a pristine comparative sample.

While the \mbox{$\Lp- F444W$} color of $\beta$~Pic~d matches those of brown dwarfs, its \mbox{$\Lp - F410M$} and \mbox{$F410M - F444W$} colors diverge significantly from those of the brown dwarfs. Recent observations of warmer planetary companions have similarly shown very faint $F410M$ magnitudes for their luminosity \citep{2025AJ....169..209B,2026ApJ..1001L..26B}. The $F444W$ filter has a pivot wavelength of 4.402\,$\mu$m and $\Delta\lambda= 1.024$\,$\mu$m, while the $F410M$ filter covers a subset of the same wavelengths, with a pivot wavelength of 4.083\,$\mu$m and $\Delta\lambda= 0.436$\,$\mu$m. $F444W$ encompasses both the 4.3\,$\mu$m CO$_2$ feature and the 4.6\,$\mu$m CO fundamental bandhead. Thus, $\beta$~Pic~d's red \mbox{$F410M - F444W$} color in particular suggests that the majority of light in the $F444W$ bandpass is emitted on the red side of the bandpass, i.e., strong CO$_2$ absorption is present on the blue side. 

This strong CO$_2$ absorption is in turn suggestive of an elevated atmospheric metallicity; we find that Sonora Flame Skimmer models (Mang et al. in prep.) with a vertical mixing parameter of $K_{zz}=4$ are able to match the \mbox{$\Lp-F410M$} colors of both the brown dwarf and planet populations, assuming that planets have higher metallicities, and the chosen $K_{zz}$ value here is consistent with that observed in previous studies of 51~Eri~b \citep{2025AJ....170..326M} and field brown dwarfs of similar temperature \citep[e.g.,][]{
2024ApJ...963...73M}. However, these models are unable to also explain the $F444W$ magnitude of $\beta$~Pic~d and other planets in the sample. 
Following \citet{2025AJ....169..209B}, who applied ExoRem models \citep{2015A&A...582A..83B, 2018ApJ...854..172C} to the spectrum of 51~Eri~b, we show the best-fitting model photometry using the same grid in Figure~\ref{fig:CMD}. This model provides a reasonable fit to all $\beta$~Pic~d photometry, explaining both the blue \mbox{$\Lp-F410M$} and red \mbox{$F410M-F444W$} colors. 

$\beta$~Pic~d has \mbox{$\Lp-F410M$} colors as well as VLT/SPHERE $K_1$-band and ERIS \Lp-band absolute magnitudes that closely match (within 1$\sigma$) 51~Eri~b \citep{2015Sci...350...64M, 2017AJ....154...10R, 2017A&A...603A..57S, 2022MNRAS.509.4411D, 2023A&A...673A..98B, 2023MNRAS.525.1375W, 2025AJ....169..209B, 2026A&A...707L..13D}. Out of the current cohort of directly imaged exoplanets, 51 Eri b is the only exoplanet with a mid-T spectral type discovered to date \citep[T6.5;][]{2017AJ....154...10R}. Like $\beta$~Pic~d, it is a BPMG member: the two planets are likely co-eval and formed from similar material. \citet{2025AJ....169..209B} recently derived \mbox{$\Teff=632\pm13$ K} and \mbox{$R=1.30\pm0.03$ \Rjup} for 51~Eri~b, based on forward-modeling of photometry and spectroscopy from Gemini/GPI, VLT/SPHERE, and JWST/NIRCam with the ExoRem model grid, confirming earlier temperature estimates from \citet{2017AJ....154...10R} and \citet{2017A&A...603A..57S}. \citet{2022MNRAS.509.4411D} derived an upper limit to the mass of 51~Eri~b of $<$9\,\Mjup based on astrometric constraints, and \citet{2023A&A...673A..98B} estimated a mass range of 2--4\,\Mjup from forward modeling of VLT/SPHERE spectroscopy.

Since 51~Eri~b is such a close analog to $\beta$~Pic~d, we estimated a bolometric correction for $\beta$~Pic~d based on the colors of 51~Eri~b following the method of Section~4.2 of \citet{2023ApJ...950L..19F}. This allows us to extrapolate a bolometric luminosity from the available photometry with a model-independent method. We adopt a bolometric correction of \mbox{BC(\Lp) = $0.06\pm0.27$ mag} from \mbox{$\log\left({\Lbol}/{L_\odot}\right) = -5.55\pm0.10$ dex} \citep{2025AJ....169..209B} and \mbox{$\Lp = 16.20\pm0.11$ mag} \citep{2017AJ....154...10R} at a distance of 29.9\,pc. Thus, for $\beta$~Pic~d, with \mbox{$M_{L^{\prime}} = 14.11\pm0.16$ mag}, our BC implies \mbox{$\log\left({\Lbol}/{L_\odot}\right) = -5.67\pm0.13$ dex}. This luminosity is consistent with a range of initial entropies, including some low enough to correspond to `cold start' formation scenarios \citep[e.g.,][]{2007ApJ...655..541M,2014MNRAS.437.1378M}. If the cold-start scenario is indeed favored for $\beta$~Pic~d, $\beta$~Pic would become the first system known to host planets formed through both hot- and cold-start pathways, with $\beta$~Pic~b and~c representing the hot start regime. However, to obtain an estimate of its mass from models, we adopt the conservative limiting case of `hot start' initial entropy.

Using this bolometric luminosity estimate for $\beta$~Pic~d with an age range of \age_myr\,Myr, we applied rejection sampling following \citet{2023MNRAS.519.1688D,2023ApJ...946L...6M,2023ApJ...951L..20C}, to estimate the physical properties of $\beta$~Pic~d. Adopting the \texttt{ATMO}\,2020 evolutionary models \citep[][both equilibrium and strong non-equilibrium grids]{2020A&A...637A..38P}, we estimate a mass of \mass_d\,\Mjup, radius of \radius_d\,\Rjup, and effective temperature of \teff_d\,K (Table~\ref{tab:exoplanet_orbit_parameters}). Considering only the ERIS \Lp-band photometry derived in this letter and interpolating directly from the same models, we find masses of 2 to 3.5\,\Mjup, consistent with the values derived through our luminosity approach, further validating our method. These values are very close matches to literature values for 51~Eri~b. Thus, $\beta$~Pic~d is the second mid-T type exoplanet discovered and is a very close analog to 51~Eri~b.

\section{Discussion}\label{sect:discussion}

\subsection{Implications for the \texorpdfstring{$\beta$~Pic system}{beta Pic system d}}

$\beta$~Pic is an extremely well-studied and dynamically active system, comprising multiple giant planets, a warped debris disk, signs of recent giant collisions, and intense exocometary activity near the star. Our discovery of an additional, massive planet in the system naturally raises several dynamical questions.

First, is $\beta$~Pic~d consistent with the disk warp? The main disk spans roughly 30--\mbox{200 au} \citep{Han2026}, with the inner \mbox{$\sim$80 au} tilted by $\sim$5\,deg relative to the outer disk \citep{1995AAS...187.3205B, Golimowski2006}. This warp was ascribed to planets b and c, whose shared orbital plane is tilted relative to the outer disk \citep{Mouillet1997, 2021A&A...654L...2L, Smallwood2023, 2026AJ....171..197M}. Our discovery of $\beta$~Pic~d is consistent with this picture; the new planet shares the orbital plane of b and c, as required by the warp model. This is further evidence that the inner and outer regions of $\beta$~Pic are generally misaligned, perhaps hinting that some past dynamical event, such as a flyby, tilted the outer system.

\begin{figure}
    \centering
    \includegraphics[width=\linewidth]{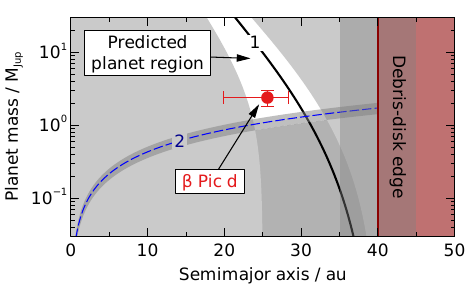}
    \caption{Dynamical prediction for a planet sculpting the disk inner edge \citep{2022A&A...659A.135P}. The predicted planet should lie in the white region to ${1\sigma}$, in agreement with our adopted parameters for $\beta$~Pic~d. Line~1 shows orbits within 5 Hill radii of the disk, along which the planet would lie, and line 2 is the minimum planet mass to sculpt the disk within the star's age. Shading around lines denotes ${1\sigma}$ uncertainties.}
    \label{fig:sculptingPlanetPrediction}
\end{figure}

Second, is $\beta$~Pic~d sculpting the disk inner edge? The other planets are too far inwards, so it has long been inferred that another planet lay beyond b and sets the disk edge through planetesimal scattering \citep[e.g.][]{2025A&A...694A.236L}. Our discovery of $\beta$~Pic~d supports this picture. The exact location of the disk edge is difficult to constrain, because the disk is viewed edge-on to Earth; a non-parametric fit to ALMA disk data yields a shallow, uncertain radial profile, whose inner edge could be anywhere from 30 to 50~au \citep{Han2026}. Assuming an inner edge of ${40\pm5}$~au, we apply the \texttt{SculptingPlanet} code\footnote{\url{https://github.com/TimDPearce/SculptingPlanet}} \citep{2022A&A...659A.135P} to dynamically predict the properties of a planet sculpting the disk edge, assuming a star mass of ${1.83\pm0.03}$\,\Msun\ and age of \age_myr\,Myr. The result is shown in \mbox{Figure \ref{fig:sculptingPlanetPrediction}}. If the disk edge is sculpted by a single planet, then that planet would lie in the white region of the figure, with a minimum mass of ${1.3 \pm 0.1}$\,\Mjup and a maximum semimajor axis of ${31 \pm 4}$\,au (higher masses and smaller semimajor axes are also compatible). This compares very favourably with our adopted parameters for $\beta$~Pic~d, with an estimated mass of \mass_d\,\Mjup and semi-major axis of \sau_d\,au. We therefore confirm that $\beta$~Pic~d is consistent with a planet sculpting the disk inner edge. Given the uncertainty on the edge location, we cannot dynamically rule out additional planets beyond d. Furthermore, the lack of significant eccentricity in the ALMA disk \citep{Lovell2026} suggests a low eccentricity for $\beta$~Pic~d. If the planet were significantly eccentric, and sufficiently massive, then it would impose an eccentricity on the disk \citep{Faramaz2014, Pearce2014}; the lack of such disk eccentricity would support an eccentricity on the lower end of our prior-dependent posterior for $e_d$ (see Section~\ref{sect:orbit_fitting}).

Third, is $\beta$~Pic~d related to the exocomets? $\beta$~Pic displays significant exocomet activity, with star-grazing comets transiting every few days \citep{Ferlet1987, LecavelierdesEtangs2022}. This likely requires planets to drive comets from a debris belt inwards towards the star (e.g.\ \citealt{Mustill2026}). One possibility is that planets b and c excite bodies from a close-in asteroid belt at ${\sim 1 \; \rm au}$ \citep{Beust2024}, whilst an alternative is that planets scatter bodies inwards from the outer belt at ${\sim 40\; \rm au}$ \citep[e.g.][]{2024MNRAS.52711664R, 2026A&A...708A..59J}. $\beta$~Pic~d could do the latter; it appears close enough to the outer belt to scatter debris, some of which could be passed inwards via planets b and c towards the star. If $\beta$~Pic~d is scattering comets inwards, then the back reaction of comets on the planet would drive $\beta$~Pic~d outwards into the belt, exposing it to fresh material and keeping the interaction going (e.g., \citealt{Bonsor2014}).

\subsection{The \texorpdfstring{$\beta$~Pic}{beta Pic} moving group as a laboratory for exoplanet formation}

The BPMG contains 5 imaged exoplanets ($\beta$~Pic~bcd, 51~Eri~b, and AF~Lep~b), as well as several free-floating planetary mass objects with similar properties to the imaged exoplanets \citep[e.g.][]{2013ApJ...777L..20L, 2021ApJ...911....7Z}. Together, these objects form a cohort of objects with a range of temperatures and masses but similar ages, both in disks around stars of varying masses and in the form of isolated planetary mass objects. This makes the BPMG a valuable laboratory for studying different modes of exoplanet formation, as well as the atmospheres and evolution of these objects \citep[e.g.][]{2011ApJ...743L..16O, 2022ApJ...934...74M, 2024A&A...687A.298N, 2025ApJ...981..138W, 2026NatAs..10..511R, 2026ApJ..1000...27X}. However, we note here that as planet formation may proceed at different rates in different systems, these planets likely do not share the exact same age \citep[e.g.][]{1996Icar..124...62P, 2007prpl.conf..607D, 2009ApJ...698....1C, 2013ApJ...776...15C}.

The colors of $\beta$~Pic~d and 51~Eri~b are notably similar. Comparative spectroscopic studies of these planets and other BPMG objects will reveal how the composition and chemistry of their atmospheres differ, and ideally also the imprint of the formation environment on these atmospheres. For the exoplanet companions, eventual dynamical mass measurements for the full sample will then allow these properties to be traced specifically as a function of mass, composition, age, and formation environment.

\subsection{A game of hide-and-seek: the necessity of long time baselines for exoplanet confirmation}

One of the main lessons of this study is that patience, as well as repeated observations, are necessary to search for and robustly confirm common proper motion for directly imaged exoplanet candidates. Due to the edge-on viewing angle of the $\beta$~Pic system, all three of the exoplanets spend significant amounts of time in undetectable positions behind the star or within the radii of coronagraphic masks. Earlier studies of the $\beta$~Pic system missed $\beta$~Pic~d largely because this planet was either at an undetectable separation from the star or very close to $\beta$~Pic~b in a given epoch of observation. This was also the case for AF~Lep~b; although this system was targeted by several imaging surveys \citep[e.g.][]{2018AJ....156..286S, 2019AJ....158...13N, 2020A&A...635A.162L}, the planet's orbital motion and high orbital inclination meant that it was not discovered until much later \citep{2023A&A...672A..94D, 2023ApJ...950L..19F, 2023A&A...672A..93M}. Planets with orbital semi-major axes $\leq$30 au which display significant orbital motion like $\beta$~Pic~d will routinely require multiple epochs to confirm. In the era of the Extremely Large Telescopes (ELTs), when direct imaging begins to observe e.g. known radial velocity exoplanets with orbital semi-major axes $<$ 10 AU, this problem will persist, as these planets may lie behind the coronagraph mask for large portions of their orbits.

Orbital motion is not the only concern here, as stars with significant proper motions of their own can in some cases mimic a common-proper motion exoplanet companion, at least for a few epochs. This has led to the retraction of multiple claims of directly imaged exoplanets, when, after a sufficiently large time baseline, it has become clear that the exoplanet candidate was in fact just a background star with significantly different proper motion than the rest of the background field \citep{2017AJ....154..218N, 2025A&A...701A.104K}. However, supplemental information can also provide compelling evidence that a candidate companion is a bonafide exoplanet before common-proper motion confirmation; in the case of $\beta$~Pic~d, its red colors (compared to the blue colors expected for background stars and galaxies), orbital motion along the plane of the known circumstellar disk, and consistency with the expected dynamical picture of the disk strongly supported that it was indeed an exoplanet.

\subsection{The potential of ELT/METIS, as revealed by ERIS-NIX}

Like JWST, ERIS-NIX opens up a new discovery space for young directly imaged exoplanets by achieving deep sensitivity at wider separations. This gain in sensitivity results from the reduced number of warm optical surfaces, made possible by mounting the instrument at the Cassegrain focus behind the deformable secondary mirror \citep{2023A&A...674A.207D}.
ERIS-NIX is equipped with two coronagraphs designed for high-contrast imaging, a vector vortex coronagraph \citep[][]{2024SPIE13097E..15O} and a vector Apodizing Phase Plate coronagraph \citep[vAPP;][]{2021ApOpt..60D..52D, 2026A&A...708A.239K}, however, the loss of throughput from coronagraphic optics limits sensitivity outside the contrast-limited regime. The ERIS-NIX non-coronagraphic imaging reported here achieved deep sensitivity, enabling a high-S/N detection of $\beta$~Pic~d in the \Lp band. This high-S/N \Lp detection then facilitated our lower S/N detections at shorter wavelengths and tighter separations. These results highlight the discovery potential of the upcoming Mid-infrared ELT Imager and Spectrograph \citep[METIS;][]{2021Msngr.182...22B} instrument for the Extremely Large Telescope (ELT), which will have both coronagraphic optics and the possibility to obtain ultra-deep, non-coronagraphic imaging with a 39-m aperture.

\section{Conclusions}\label{sect:conclusions}

We present the discovery of $\beta$~Pic~d, a third, widely-separated giant planet in the $\beta$~Pictoris system, with detections in 6 epochs of observations covering a total baseline of 11~years. Our findings are summarised below:

\begin{itemize}
  \item We detect $\beta$~Pic~d at a contrast of $\Delta\Lp=12.11\pm0.15$\,mag and separation of 1126.9$\pm$3.9~mas in our VLT/ERIS observations from 2025~December~3 and at closer separations in earlier JWST/NIRCam and VLT/SPHERE observations in several filters. Our astrometry shows that its motion is consistent with a gravitationally-bound source with orbital motion approximately in the plane of the $\beta$~Pic disk, and rules out the possibility that it is a background star or galaxy. $\beta$~Pic~d is currently moving outwards in its orbit, and was likely obscured by its host star and $\beta$~Pic~b in archival VLT/NaCo observations from $\sim$2009 to 2011.

  \item We estimate a mass of \mass_d\,\Mjup, radius of \radius_d\,\Rjup, and \Teff\ of \teff_d\,K for $\beta$~Pic~d, interpolating from the \texttt{ATMO}\,2020 evolutionary models. Its colors differ from brown dwarfs at similar absolute magnitudes and its red $F410M - F444W$ color suggests the presence of CO$_2$ in its atmosphere.

  \item We use astrometry, along with literature measurements of the inner planets and host star, to fit the orbit and measure an inclination and longitude of the ascending node that are consistent with being coplanar with the inner planets. After excluding orbits that cross $\beta$~Pic~b's orbit, we find a semi-major axis of \sau_d\,au, which corresponds to a period of \per_d\,yr, and eccentricity $<0.44$ (2$\sigma$). Such orbits are consistent with theoretical predictions for a planet sculpting the debris-disk inner edge.
  
  \item $\beta$~Pic~d is a close analog of $\beta$~Pic moving group exoplanet 51~Eri~b and is only the second exoplanet discovered with a likely mid-T spectral type. This makes the $\beta$~Pic planetary system the first known to contain both definitively `hot-start' planets (b and~c) and a potentially `cold-start' planet. $\beta$~Pic~d is the least massive exoplanet detected in the $\beta$~Pic system to date, and is among the lowest-mass exoplanets imaged from the ground.
\end{itemize}

The discovery of $\beta$~Pic~d using a combination of ground and space-based observations demonstrates the complementarity of these facilities and the value of archival datasets for investigating and confirming newly identified companions, even in well-known systems. $\beta$~Pic~d was missed in earlier studies of the system, in part because the planet was too close to its host star or the far brighter $\beta$~Pic~b for many epochs of observation. This illustrates the importance of revisiting compelling targets when conducting planet searches. Furthermore, the strong \Lp-band detection of this planet with VLT/ERIS emphasises the deep contrast sensitivity that large ground-based telescopes still offer in the era of JWST, and hints at the potential of ELT/METIS for detecting even fainter, lower-mass exoplanets at wavelengths closer to the peak of their emission. 

Future observations of $\beta$~Pic~d will characterise its atmosphere in depth, further constrain its orbit, and yield deeper insights into the architecture and formation of this young planetary system. This discovery also highlights the value of the BPMG, which now contains 5 imaged exoplanets with a range of temperatures and masses but similar ages, as a laboratory for understanding planetary atmospheres and evolution. Spectroscopic studies and dynamical mass measurements of this sample will probe how the properties of these planets differ, and thus reveal the imprint of their formation environment on their atmospheres.

\section*{Data Availability}

A data package for this article will be made available through a Zenodo repository shortly after publication, at \dataset[10.5281/zenodo.20246422]{https://doi.org/10.5281/zenodo.20246422}. This will include either the datasets themselves or links to the archives from which they can be downloaded, as well as the config files for our orbit fitting procedure. The JWST/NIRCam data presented in this article were obtained from the Mikulski Archive for Space Telescopes (MAST) at the Space Telescope Science Institute. The specific observations analyzed can be accessed via \dataset[doi:10.17909/wk9r-7797]{https://doi.org/10.17909/wk9r-7797}.

\begin{acknowledgments}
The authors would like to thank the referee of this letter, Thayne Currie, whose thoughtful and constructive comments helped us to improve this work.

The authors thank the JWST GTO~1411 (PI: Stark) and GO~4758 (PI: Zhou) teams for providing reduced data products and valuable advice. 
BJS and BAB acknowledge funding by the UK Science and Technology Facilities Council (STFC) grant nos. ST/V000594/1 and UKRI1196.
LTP and JLB acknowledge funding from the European Research Council (ERC) under the European Union’s Horizon 2020 research and innovation program under grant agreement No 805445.
LTP further acknowledges funding support from Brasenose College.
JLB further acknowledges the support of the Leverhulme Trust via the Philip Leverhulme Physics Prize.
TDP is supported by a UKRI Stephen Hawking Fellowship and a Warwick Prize Fellowship, the latter made possible by a generous philanthropic donation.
EOG gratefully acknowledges the financial support from the Swiss National Science Foundation (SNSF) under project grant number 200020\textunderscore200399.
LI is supported by funding from Breakthrough Listen. The Breakthrough Prize Foundation funds the Breakthrough Initiatives, which manages Breakthrough Listen.
SH acknowledges funding by the UK Science and Technology Facilities Council (STFC) grant no. ST/Y002792/1.
This work has been carried out within the framework of the NCCR PlanetS supported by the Swiss National Science Foundation under grants 51NF40\textunderscore182901 and 51NF40\textunderscore205606. 
This project has received funding from the European Research Council (ERC) under the European Union’s Horizon 2020 research and innovation program (COBREX; grant agreement \#885593).
JMV acknowledges support from a Royal Society - Research Ireland University Research Fellowship (URF/1/221932, RF/ERE/221108) and the European Union through the Exo-PEA ERC project (grant number 101164652). Views and opinions expressed are however those of the author(s) only and do not necessarily reflect those of the European Union or the European Research Council Executive Agency. Neither the European Union nor the granting authority can be held responsible for them.
This material is based upon work supported by the National Aeronautics and Space Administration under Agreement No. 80NSSC21K0593 for the program ``Alien Earths.’'
ECG acknowledges support from the Heising-Simons foundation.

This work is based on observations collected at the European Organisation for Astronomical Research in the Southern Hemisphere under ESO programmes 116.28WQ.001, 60.A-9255(A), 60.A-9382(A), and 0104.C-0418(F). We would like to thank the support staff at Paranal Observatory for their assistance in obtaining these observations. This work is based in part on observations made with the NASA/ESA/CSA JWST. The data were obtained from the Mikulski Archive for Space Telescopes at the Space Telescope Science Institute, which is operated by the Association of Universities for Research in Astronomy, Inc., under NASA contract NAS 5-03127 for JWST. These observations are associated with JWST programs GTO~1411 (PI: Stark) and GO~4758 (PI: Zhou).

This research has made use of the NASA Exoplanet Archive, which is operated by the California Institute of Technology, under contract with the National Aeronautics and Space Administration under the Exoplanet Exploration Program \citep{2025PSJ.....6..186C}. This research has made use of the Astrophysics Data System, funded by NASA under Cooperative Agreement 80NSSC21M00561. This research has made use of the SIMBAD database, operated at CDS, Strasbourg, France \citep{2000A&AS..143....9W}. This research made use of SAOImageDS9, a tool for data visualization supported by the Chandra X-ray Science Center (CXC) and the High Energy Astrophysics Science Archive Center (HEASARC) with support from the JWST Mission office at the Space Telescope Science Institute for 3D visualization \citep{2003ASPC..295..489J}. This work has made use of data from the European Space Agency (ESA) mission {\it Gaia} (\url{https://www.cosmos.esa.int/gaia}), processed by the {\it Gaia} Data Processing and Analysis Consortium (DPAC, \url{https://www.cosmos.esa.int/web/gaia/dpac/consortium}) \citep[][]{2016A&A...595A...1G, 2023A&A...674A...1G}. Funding for the DPAC has been provided by national institutions, in particular the institutions participating in the {\it Gaia} Multilateral Agreement. This research has made use of the Spanish Virtual Observatory (SVO) Filter Profile Service ``Carlos Rodrigo", funded by MCIN/AEI/10.13039/501100011033/ through grant PID2023-146210NB-I00 \citep{2020sea..confE.182R}. This work made use of the whereistheplanet\footnote{\url{http://whereistheplanet.com/}} prediction tool \citep{2021ascl.soft01003W} to initialize the removal of $\beta$~Pic~b for our NegFC analysis. This work has made use of the \texttt{species} package for synthetic photometry and mass estimates using model isochrones \citep{2020A&A...635A.182S}.
\end{acknowledgments}

\begin{contribution}

B.J.S. and M.J.B. were equally responsible for leading this project as well as the preparation and presentation of this manuscript. B.J.S. led the coordination of the team. B.J.S. also led the analysis of the JWST/NIRCam data, and M.J.B. led the analysis of the VLT/ERIS and VLT/NaCo data. V.C. led the analysis of the VLT/SPHERE data with help of M.J.B. for the preliminary orbit fits. C.F. and T.D. led the orbit fitting analysis. E.C.M. and B.A.B. led the interpretation of the photometry of $\beta$~Pic~d. L.T.P. and B.J.S. acquired the VLT/ERIS observations. T.D.P. led the analysis of the implications of this discovery for the $\beta$~Pictoris disk and system as a whole. J.L.B. and B.A.B. led the UK ERIS GTO team and were the PIs of the VLT/ERIS dataset that led to the discovery of $\beta$~Pic~d. Authors E.O.G. through B.S. provided relevant scientific expertise and contributed to the preparation of the manuscript. All other authors (G.A. through M.X.) are listed alphabetically and made crucial contributions to acquiring the observations described in this letter or designing and developing the VLT/ERIS instrument central to this work.

\end{contribution}

%
\facilities{VLT:Yepun (ERIS/NIX, NaCo), VLT:Melipal (SPHERE), VLT:Antu (NaCo), JWST}

\software{This work makes use of the Python programming language\footnote{Python Software Foundation; \url{https://www.python.org/}}, in particular packages including \texttt{NumPy} \citep{harris2020array}, \texttt{backtracks} \citep{backtracks_code}, \texttt{Octofitter} \citep{Thompson_2023}, \texttt{orbitize} \citep{2020AJ....159...89B, 2024JOSS....9.6756B}, \texttt{orvara} \citep{2021AJ....162..186B}, \texttt{Astropy} \citep{2013A&A...558A..33A, 2018AJ....156..123A, 2022ApJ...935..167A}, \texttt{astroquery} \citep{2019AJ....157...98G}, \texttt{esorex} \citep{2015ascl.soft04003E}, \texttt{pyKLIP} \citep{2015ascl.soft06001W}, \texttt{spaceKLIP} \citep{2022SPIE12180E..3QG, 2022SPIE12180E..3NK, 2023ApJ...951L..20C}, \texttt{STPSF} \citep{2012SPIE.8442E..3DP, 2014SPIE.9143E..3XP}, \texttt{fourS} \citep{2025AJ....169..194B}, \texttt{eristools} (Bonse et al. in prep.), \texttt{PynPoint} \citep{2012MNRAS.427..948A, 2019A&A...621A..59S}, \texttt{vcal-sphere} \citep{Christiaens2023b}, \texttt{vip\_hci} \citep{GomezGonzalez2017, Christiaens2023}, \texttt{SculptingPlanet} \citep{2022A&A...659A.135P}, \texttt{species} \citep{2020A&A...635A.182S}, \texttt{Matplotlib} \citep{Hunter:2007}, \texttt{SciPy} \citep{2020SciPy-NMeth}, \texttt{emcee} \citep{2013PASP..125..306F}, and \texttt{uncertainties} \citep{eric_o_lebigot_eol_2025_15257927}.}


\appendix
\section{Non-detections}
\label{sect:non_detections}
\begin{deluxetable*}{l c c c r c c l}
\tabletypesize{\scriptsize}
\tablecaption{List of $\beta$~Pic observations analysed here that did not yield a detection. \label{tab:obs_non}}
\tablecolumns{6}
\tablehead{
\colhead{Instrument}  & \colhead{Program}  & \colhead{Obs. Date} & \colhead{Filter}  & \colhead{Exp.} & \colhead{Seeing}   & \colhead{Field rot.} & \colhead{Comment}\\
\colhead{}            & \colhead{}            & \colhead{(YMD-UT)}  & \colhead{}        & \colhead{(s)}  & \colhead{(\arcsec)} & \colhead{(deg)} & \colhead{}
}
\startdata
JWST/NIRCam  & GO~4758 & 2025-03-22 & $F210M$  & 47\,040  & \nodata    & 9.9   \\
JWST/NIRCam  & GO~4758 & 2023-03-18 & $F182M$  & 3\,014   & \nodata    & 10.2  \\
JWST/NIRCam  & GTO~1411 & 2023-03-18 & $F210M$  & 3\,014   & \nodata    & 10.2  \\
JWST/NIRCam  & GTO~1411 & 2023-03-18 & $F250M$  & 3\,421   & \nodata    & 10.2  \\
JWST/NIRCam  & GTO~1411 & 2023-03-18 & $F300M$  & 3\,421   & \nodata    & 10.2  \\
JWST/NIRCam  & GTO~1411 & 2023-03-18 & $F335M$  & 3\,667   & \nodata    & 10.2  \\
\hline
VLT/SPHERE  & 0104.C-0418(F)  & 2020-02-08 & $YJ$    & 2\,024   & 0.60          & 30.7 & IFS dataset.\\
VLT/SPHERE  & 0104.C-0418(D)  & 2019-12-22 & $H23$    & 1\,260   & 0.59          & 30.7 & Short total integration. \\
VLT/SPHERE  & 60.A-9382(A)  & 2019-03-11 & $YJH$    & 2\,808   & 0.72          & 32.1 & IFS dataset.\\
VLT/SPHERE   & 1100.C-0481(V) & 2018-09-12 & $H23$    & 1\,144   & 0.57          & 26.1 & Short total integration and high airmass ($\sim$1.5). \\
VLT/SPHERE & 0104.C-0418(F) & 2016-10-14 & $H23$    & 3\,872   & 0.72       & 53.5 & Wind-driven halo. Tentative detection of d. \\
VLT/SPHERE & 096.C-0241(E)   & 2016-03-26 & $H23$    & 1\,664 & 1.65  & 19.9\\
VLT/SPHERE & 096.C-0241(G)   & 2016-01-20 & $H23$    & 2\,328  & 1.15  & 29.8 & \\
VLT/SPHERE & 096.C-0241(B)  & 2015-11-28 & $H23$    & 2\,104   & 1.13  & 40.0 & Wind-driven halo.\\
VLT/SPHERE  & 095.C-0298(H)  & 2015-02-05 & $H23$    & 3\,384        & 0.75 & 43.3 & Very close proximity to b. Tentative detection of d.\\
\hline
VLT/NaCo    & 094.C-0149(D) & 2015-01-23 & \Lp  & 4699  & 0.99      & 30.6  & Poor seeing and little field rotation.\\
VLT/NaCo    & 090.C-0396(D) & 2012-12-16 & \Lp  & 3360  & 0.72      & 67.6  & $\beta$~Pic~d is too close to the star ($<300$~mas)\\
VLT/NaCo    & 088.C-0358(C) & 2011-12-11 & \Lp  & 4800  & 0.57      & 54.1  & $\beta$~Pic~d is too close to the star ($<300$~mas)\\
VLT/NaCo    & 086.C-0341(B) & 2010-09-28 & \Lp  & 6000  & 0.80      & 51.7  & $\beta$~Pic~d is too close to the star ($<300$~mas)\\
VLT/NaCo    & 284.C-5057(A) & 2010-04-07 & \Lp  & 3360  & 1.23      & 18.3  & $\beta$~Pic~d is too close to the star ($<300$~mas)\\
VLT/NaCo    & 284.C-5057(A) & 2010-03-21 & \Lp  & 3360  & 1.36      & 19.9  & $\beta$~Pic~d is too close to the star ($<300$~mas)\\
VLT/NaCo    & 084.C-0739(A) & 2009-12-29 & \Lp  & 6060  & 0.83      & 61.0  & $\beta$~Pic~d is too close to the star ($<300$~mas)\\
VLT/NaCo    & 084.C-0739(A) & 2009-12-26 & \Lp  & 4800  & 0.86      & 44.0  & $\beta$~Pic~d is too close to the star ($<300$~mas)\\
VLT/NaCo    & 084.C-0739(A) & 2009-12-25 & \Lp  & 4800  & 1.18      & 50.8  & $\beta$~Pic~d is too close to the star ($<300$~mas)\\
VLT/NaCo    & 072.C-0624(B) & 2003-11    & \Lp  & 4113  & 0.69      & \nodata & No pupil-tracking. \\
\enddata
\tablecomments{
 The VLT/NaCo data taken in 2003-11 were collected in field-tracking between 2003-11-10 and 2003-11-17. Additional frames of the reference star \object{HR~2435} were used for RDI.
}
\end{deluxetable*}
In addition to the detection in the archival data, we processed several datasets that did not result in the detection of $\beta$~Pic~d. These non-detections are attributable to the planet being too close to the star or to poor observing conditions. Despite all SPHERE data being acquired in dual IRDIS+IFS modes, we only reduced the 2019 and 2020 IFS datasets corresponding to the IRDIS data in which we did detect $\beta$~Pic~d. This is for two reasons. Firstly, 16 out of 18 datasets were taken in IFS ($YJ$) + IRDIS ($H23$) mode, where the IFS data span a short wavelength coverage, which limits the efficiency of SDI,  where planet d is expected to be extremely faint. This includes the 2020 IFS data. Secondly, among the two remaining SPHERE datasets, acquired in IFS ($YJH$) + IRDIS ($K12$) mode (i.e., the 2014 and 2019 epochs), planet d was found at a much shorter separation and in the wings of planet~b in the 2014 IRDIS data. This leaves the 2019 $YJH$ IFS dataset as the most amenable for a re-detection of planet d at shorter wavelengths. All the investigated datasets are summarized in Table~\ref{tab:obs_non}.

\section{Subtraction of \texorpdfstring{$\beta$~Pic~b}{beta Pic b} in the 2014 VLT/SPHERE data}
\label{sect:SubtractionOfb}
In 2014, $\beta$~Pic~d was very close to $\beta$~Pic~b, which outshines the faint companion by several orders of magnitude. To detect and characterize $\beta$~Pic~d, we used the NegFC technique implemented in VIP \citep[][]{GomezGonzalez2017, Christiaens2023}. 
We used NegFC twice: first, to model and subtract $\beta$~Pic~b to enable the detection of $\beta$~Pic~d, and second, to fit the astrometry and photometry of $\beta$~Pic~d. The analysis was carried out separately for the $K1$ and $K2$ bands.

\subsection{Detection of \texorpdfstring{$\beta$~Pic~d}{beta Pic d} in 2014}
\begin{figure}
    \includegraphics[width=\linewidth]{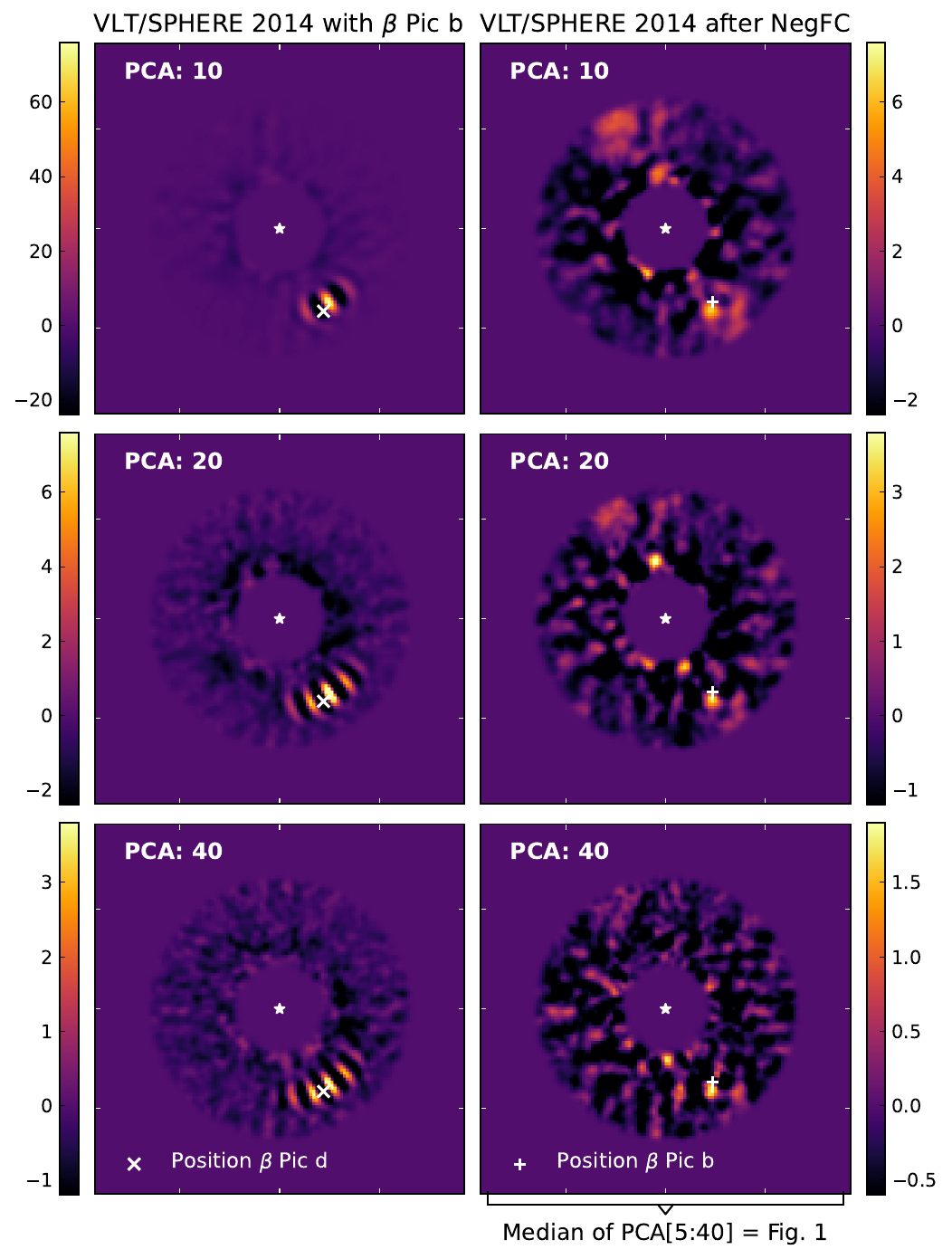}
    \caption{PSF-subtracted images obtained with annular PCA on the original 2014 VLT/SPHERE datacube (left column) and the cube resulting from the subtraction of our best estimate of $\beta$~Pic~b (right column), shown for 10, 20 and 40 principal components. The residuals shown are the results for the $K1$ filter. The ticks are spaced by 500~mas intervals.}
    \label{fig:SubtractionOfPlanetb}
\end{figure}
Unlike explicit forward models \citep[e.g.,][]{2016ApJ...824..117P}, which estimate the signature of a companion in the PCA residuals, the NegFC technique inserts a negative fake companion into the raw frames before PCA. This approach achieves a better accuracy than forward models to recover the astrometry and photometry of bright companions \citep[e.g.][]{Cantalloube2024}. We refer to the data after the fake companion is subtracted as the \emph{empty cube}.
The empty cube is treated as a new dataset; that is, a new PCA basis is calculated to obtain the residual image. 
Using Nelder–Mead minimization, we optimized the position and flux of the inserted negative companion by minimizing an objective function defined as the standard deviation of the residual pixel intensities within a 1-FWHM aperture centred on the first-guess location of $\beta$~Pic~b.
This process revealed $\beta$~Pic~d next to the original position of $\beta$~Pic~b (see Figure~\ref{fig:SubtractionOfPlanetb}, for residuals in the $K1$ filter).
We assessed the robustness of this detection by applying annular PCA with 1 to 100 components on the optimal empty cube, i.e., the cube that minimizes the NegFC objective function.
The dataset offers a total field rotation of $26.1$~deg, resulting in rotational motion of $171$~mas. To preserve as many frames as possible for accurate subtraction of the speckle noise with PCA, we did not apply a rotation threshold.
While the speckle pattern changes as a function of the principal components, the detection of $\beta$~Pic~d is robust for a wide range of components (5 to 40).
The left column of Figure~\ref{fig:SubtractionOfPlanetb} shows residual images calculated on the original ADI cubes, including $\beta$~Pic~b.
As the number of principal components increases, the signal of $\beta$~Pic~b is altered,  resulting in a stripe pattern. This effect is caused by the brightness of the companion, which directly influences the PCA basis. 
The robust recovery of $\beta$~Pic~d for a wide range of principal components confirms that it is not a residual from a poor subtraction of $\beta$~Pic~b.

The residual image shown in Figure~\ref{fig:final_images} was obtained by median combining all residual images between five and 40 components, and then averaging the result over the $K1$ and $K2$ filters.

\subsection{Astrometry and photometry of \texorpdfstring{$\beta$~Pic~d}{beta Pic d}}\label{sect:astrometry_photometry_2014}
As for the other epochs, we estimated the astrometry and photometry of $\beta$~Pic~d using NegFC, but this time on the empty cube after subtracting $\beta$~Pic~b.
To propagate the uncertainties associated with the subtraction of $\beta$~Pic~b into the retrieved parameters of $\beta$~Pic~d, we used the MCMC implementation of NegFC in VIP to draw samples from the posterior distribution of the position and flux of $\beta$~Pic~b.
For each posterior sample of the $\beta$~Pic~b parameters, we generated a new empty cube and repeated the NegFC inference for $\beta$~Pic~d. We performed this procedure for 30 randomly selected posterior samples. For each realization, we considered two aperture radii for the objective function, 1~FWHM and 2~FWHM, and six 20-component PCA ranges between 10--30, and 60--80 components. Within each range, the corresponding PCA residuals were median-combined before evaluating the NegFC objective function. 
These choices reflect the facts that the speckle noise and residuals from $\beta$~Pic~b vary depending on the number of principal components, and that either choice of aperture size yielded plausible residual-minimization solutions.
Through this analysis, we found that the uncertainty associated with the subtraction of $\beta$~Pic~b dominates the error budget of the inferred $\beta$~Pic~d parameters. By contrast, injection-and-recovery tests of individual fake planets with the inferred $\beta$~Pic~d parameters, but placed at different azimuths, yielded negligible additional uncertainties.

The final adopted uncertainties reflect the spread in the retrieved $\beta$~Pic~d parameters induced by varying the PCA range, the aperture radius used for the NegFC objective function, and the posterior sample of the $\beta$~Pic~b parameters used to construct the empty cube.
We estimated contrast uncertainties for $\beta$~Pic~d of $\sim$0.7~mag and $\sim$0.6~mag in the $K1$ and $K2$ bands, respectively. 
Our final astrometric uncertainties reflect the spread of radial separation and PA values inferred for both $K1$ and $K2$ (Table~\ref{tab:obs}).

\section{Orbit Analysis}
\label{A:orbit-fit}

We performed an analysis of $\beta$~Pic~d's orbit using our new relative astrometry as well as data from the literature. We used \texttt{orvara} \citep[v1.1.2;][]{2021AJ....162..186B}, which can fit relative astrometry for multiple planets, host-star RVs, and absolute astrometry from \textsl{Hipparcos} and \textsl{Gaia}~DR3. It accounts for the astrometric offsets of any planets interior to a given planet and neglects the influence of outer planets, which has been shown to be valid for the orbits of $\beta$~Pic~c and $\beta$~Pic~b to $\approx$20\,$\mu$as over 20~years \citep{2021AJ....161..179B}. The same work also showed that the impact of ignoring $\beta$~Pic~c when fitting $\beta$~Pic~b astrometry led to systematics of $\sim$10\,mas. Given that $\beta$~Pic~b's orbit is about 4$\times$ larger than $\beta$~Pic~c, its influence on $\beta$~Pic~d's astrometry could be even larger, although the timescale of orbit variations is correspondingly longer, which may reduce its influence.

In the following, we adopted priors that are uniform in $\log{M_{\star}}$ and $\log{\sigma_{\rm jit}}$ for the host star and $\log{a}$, $e$, $\Omega$, $\omega$, $\lambda$, and $\sin{i}$, for all planets. To improve convergence, we used a start file to initialize our chains at values near the highest-likelihood parameters found in preliminary exploratory chains, and we imposed loose Gaussian priors on the masses of planets~b and~c of $0.009\pm0.018$\,\Msun and $0.008\pm0.016$\,\Msun, respectively. Mainly due to the uncertain nature of the mass of $\beta$~Pic~d, we chose to neglect its mass in our orbit fits. By construction, it would not influence \texttt{orvara}'s fit of the inner two planets at all, and it is expected to be much smaller in mass than the inner planets, so it should induce a comparatively negligible acceleration on the host star ($\propto m_{\rm d}/r_{\rm d}^2$). We used 100 walkers, five parallel-tempering temperatures, $10^6$ steps, thinned every 50th step, and we discarded the first 50\% of steps as burn-in. The configuration files used for our orbit fit are provided in the Zenodo data package connected to this manuscript.

\subsection{Literature data}
\label{A:lit_data}

To perform our multi-planet orbit analysis, we included literature measurements of relative astrometry and RVs, as well as the most recent \textsl{Hipparcos}--\textsl{Gaia} Catalog of Accelerations \citep[HGCA;][]{2021ApJS..254...42B}. We used the host-star RVs from \citet{2019NatAs...3.1135L} cleaned of stellar activity as presented by \citet{2020AJ....160..243V}. For the innermost planet $\beta$~Pic~c, only VLTI/GRAVITY has resolved it, and we used all four reprocessed measurements from \citet{2021A&A...654L...2L}. For $\beta$~Pic~b, we use the following: all nine VLT/NACO measurements from 2003--2011 processed by \citet{2012A&A...542A..41C}; the 2008 $M^{\prime}$-band detection from \citet{2011ApJ...736L..33C}; the 2012 epoch from \citet{2013A&A...555A.107B}; eight Gemini/NICI and Magellan/MagAO measurements over 2009--2012 from \citet[][which excludes one discrepant $CH_4S$ measurement from 2012]{2014ApJ...794..158N}; fourteen Gemini/GPI measurements spanning 2013--2018 from \citet{2016AJ....152...97W} as reported in \citet[][which excludes one 2013 measurement with an anomalously high PA]{2020AJ....159...71N}; all 34 VLT/SPHERE measurements spanning 2014--2020 from \citet{Lagrange2019} and \citet{2020A&A...642A..18L}; and all seven VLT/GRAVITY measurements from 2018--2021 as reprocessed by \citet{2021A&A...654L...2L}. The HGCA does not contain a significant detection of astrometric acceleration ($\chi^2=0.935$), but it can still be useful in constraining the planet masses as the \textsl{Hipparcos} and \textsl{Gaia} proper motion anomaly nondetections have uncertainties of about 0.4\,mas\,yr$^{-2}$ and 0.3\,mas\,yr$^{-2}$, respectively, so we include it.

\begin{figure*}
    \includegraphics[width=\textwidth]{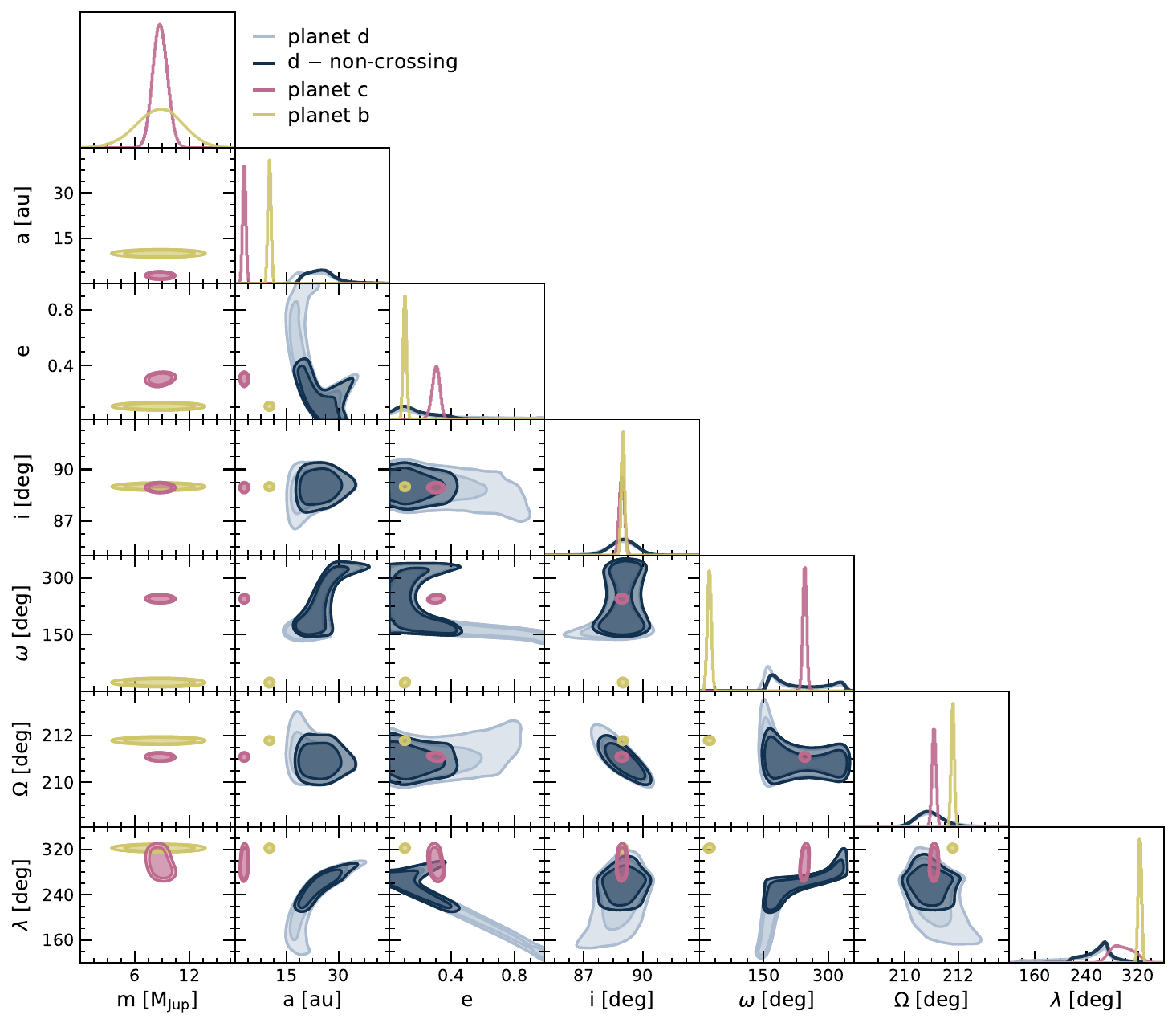}
    \caption{Marginalised posterior distributions for our orbital parameters of planets~d (grey), c~(red), and b~(yellow) in the $\beta$~Pic system. Contour lines and shaded regions in the joint posterior indicate the 1-$\sigma$ and 2-$\sigma$ credible regions. The posteriors for planet~d include the full posterior distributions from the fits (light grey) along with the posteriors after removal of solutions crossing the orbits of the inner planets (dark grey). As planet~d was set to be mass-less, it is not shown in the first column. The orbital parameters shown are the semi-major axes, eccentricities, inclinations, arguments of periapsis $\omega$, longitudes of ascending node $\Omega$ and mean longitudes $\lambda$ at reference epoch $T_\mathrm{ref}=2010.0$.}
    \label{fig:full_corner}
\end{figure*}

\begin{deluxetable*}{lccc}
\tablecaption{Posteriors from \texttt{orvara} analysis \label{tbl:orvara}}
\setlength{\tabcolsep}{0.10in}
\tabletypesize{\normalsize}
\tablewidth{0pt}
\tablehead{
\colhead{Parameter}             &
                                &
\colhead{Mode $\pm$1$\sigma$} &
\colhead{95.4\% c.i.}           }
\startdata
$M_{\star}$ (\Msun)              & * & $1.789$\raisebox{0.5ex}{\tiny$\substack{+0.024 \\ -0.027}$}      &     $1.737$, $1.838$              \\
$\sigma_{\rm jit}$ (m\,s$^{-1}$) & * & $48$\raisebox{0.5ex}{\tiny$\substack{+10 \\ -8\phn}$}            &     $33$, $70$                    \\
$m_{\rm b}$ (\Mjup)              & * & $8.7$\raisebox{0.5ex}{\tiny$\substack{+2.6 \\ -2.5}$}            & \phn$3.5$, $13.7$                 \\
$a_{\rm b}$ (au)                 & * & $10.07\pm0.03$                                                   &     $10.00$, $10.13$              \\
$e\sin\omega_{\rm b}$            & * & $0.131$\raisebox{0.5ex}{\tiny$\substack{+0.017 \\ -0.016}$}      &     $0.097$, $0.164$              \\
$e\cos\omega_{\rm b}$            & * & $0.2966$\raisebox{0.5ex}{\tiny$\substack{+0.0023 \\ -0.0022}$}   &     $0.2919$, $0.3010$            \\
$e_{\rm b}$                      &   & $0.105$\raisebox{0.5ex}{\tiny$\substack{+0.004 \\ -0.003}$}      &     $0.098$, $0.113$              \\
$\omega_{\rm b}$ (deg)           &   & $23.7$\raisebox{0.5ex}{\tiny$\substack{+3.1 \\ -2.6}$}           &     $17.9$, $29.3$                \\
$i_{\rm b}$ (deg)                & * & $88.993$\raisebox{0.5ex}{\tiny$\substack{+0.010 \\ -0.011}$}     &     $88.972$, $89.013$            \\
$\Omega_{\rm b}$ (deg)           & * & $211.787$\raisebox{0.5ex}{\tiny$\substack{+0.010 \\ -0.008}$}\phn&     $211.771$, $211.806$          \\
$\lambda_{\rm b}$ (deg)          & * & $322.6$\raisebox{0.5ex}{\tiny$\substack{+0.8 \\ -0.6}$}\phn      &     $321.2$, $324.1$              \\
$T^{\rm peri}_{\rm b}$ (yr)      &   & $1990.27\pm0.04$\phn                                             &     $1990.19$, $1990.35$          \\
$P_{\rm b}$ (yr)                 &   & $23.77$\raisebox{0.5ex}{\tiny$\substack{+0.15 \\ -0.16}$}        &     $23.45$, $24.08$              \\
$m_{\rm c}$ (\Mjup)              & * & $8.7$\raisebox{0.5ex}{\tiny$\substack{+0.8 \\ -0.7}$}            & \phn$7.2$, $10.4$                 \\
$a_{\rm c}$ (au)                 & * & \phn$2.70\pm0.03$                                                &     $2.65$, $2.76$                \\
$e\sin\omega_{\rm c}$            & * & $-0.504$\raisebox{0.5ex}{\tiny$\substack{+0.026 \\ -0.022}$}\phs &     $-0.545$, $-0.444$            \\
$e\cos\omega_{\rm c}$            & * & $-0.237$\raisebox{0.5ex}{\tiny$\substack{+0.024 \\ -0.016}$}\phs &     $-0.268$, $-0.191$            \\
$e_{\rm c}$                      &   & $0.307$\raisebox{0.5ex}{\tiny$\substack{+0.019 \\ -0.025}$}      &     $0.254$, $0.346$              \\
$\omega_{\rm c}$ (deg)           &   & $-114.9$\raisebox{0.5ex}{\tiny$\substack{+2.6 \\ -2.4}$}\phs\phn &     $-120.3$, $-109.9$            \\
$i_{\rm c}$ (deg)                & * & $88.93$\raisebox{0.5ex}{\tiny$\substack{+0.11 \\ -0.09}$}        &     $88.74$, $89.14$              \\
$\Omega_{\rm c}$ (deg)           & * & $211.08$\raisebox{0.5ex}{\tiny$\substack{+0.05 \\ -0.04}$}\phn   &     $211.00$, $211.18$            \\
$\lambda_{\rm c}$ (deg)          & * & $288$\raisebox{0.5ex}{\tiny$\substack{+23 \\ -13}$}              &     $266$, $329$                  \\
$T^{\rm peri}_{\rm c}$ (yr)      &   & $2019.490$\raisebox{0.5ex}{\tiny$\substack{+0.018 \\ -0.027}$}\phn\phn\phn&$2019.429$, $2019.525$        \\
$P_{\rm c}$ (yr)                 &   & $3.29$\raisebox{0.5ex}{\tiny$\substack{+0.07 \\ -0.03}$}         &     $3.23$, $3.42$                \\
$a_{\rm d}$ (au)                 & * & \nodata                                                          &     $14.9$, $36.7$                \\
$e\sin\omega_{\rm d}$            & * & \nodata                                                          &     $-0.44$, $0.52$\phs           \\
$e\cos\omega_{\rm d}$            & * & $-0.80$\raisebox{0.5ex}{\tiny$\substack{+0.69 \\ -0.04}$}\phs    &     $-0.85$, $0.48$\phs           \\
$e_{\rm d}$                      &   & $0.10$\raisebox{0.5ex}{\tiny$\substack{+0.31 \\ -0.10}$}         &     $0.00$, $0.90$                \\
$\omega_{\rm d}$ (deg)           &   & \nodata                                                          &     $-170$, $180$\phs             \\
$i_{\rm d}$ (deg)                & * & $88.9$\raisebox{0.5ex}{\tiny$\substack{+0.9 \\ -0.7}$}           &     $86.5$, $90.7$                \\
$\Omega_{\rm d}$ (deg)           & * & $210.9$\raisebox{0.5ex}{\tiny$\substack{+0.7 \\ -0.6}$}\phn      &     $209.6$, $213.5$              \\
$\lambda_{\rm d}$ (deg)          & * & $270$\raisebox{0.5ex}{\tiny$\substack{+17 \\ -64}$}              &     $135$, $299$                  \\
$T^{\rm peri}_{\rm d}$ (yr)      &   & $2003\pm10$                                                      &     $1985$, $2030$                \\
$P_{\rm d}$ (yr)                 &   & \nodata                                                          & \phn$42$, $165$                   \\
\enddata
\tablecomments{A ``*'' next to a parameter indicates that it was directly varied in the MCMC; any other parameters were computed from combinations of these. For parameters lacking their mode $\pm$1$\sigma$ values, the posterior was strongly multimodal. The reference epoch is 2010.0.}
\end{deluxetable*}

\begin{deluxetable*}{lccc}
\tablecaption{Posteriors for $\beta$~Pic~d excluding orbits that cross $\beta$~Pic~b's orbit \label{tbl:orvara-no-X}}
\setlength{\tabcolsep}{0.10in}
\tabletypesize{\normalsize}
\tablewidth{0pt}
\tablehead{
\colhead{Parameter}             &
                                &
\colhead{Mode $\pm$1$\sigma$} &
\colhead{95.4\% c.i.}           }
\startdata
$a_{\rm d}$ (au)               & * & $26.0$\raisebox{0.5ex}{\tiny$\substack{+2.2 \\ -6.1}$}           &     $17.9$, $39.7$                \\
$e\sin\omega_{\rm d}$          & * & $-0.26$\raisebox{0.5ex}{\tiny$\substack{+0.30 \\ -0.09}$}\phs    &     $-0.42$, $0.21$\phs           \\
$e\cos\omega_{\rm d}$          & * & $-0.57$\raisebox{0.5ex}{\tiny$\substack{+0.62 \\ -0.05}$}\phs    &     $-0.64$, $0.54$\phs            \\
$e_{\rm d}$                    &   & $0.10$\raisebox{0.5ex}{\tiny$\substack{+0.12 \\ -0.10}$}         &     $0.00$, $0.44$                \\
$\omega_{\rm d}$ (deg)         &   & \nodata                                                          &     $-170$, $180$\phs             \\
$i_{\rm d}$ (deg)              & * & $89.0$\raisebox{0.5ex}{\tiny$\substack{+0.7 \\ -0.6}$}           &     $87.7$, $90.3$                \\
$\Omega_{\rm d}$ (deg)         & * & $210.8$\raisebox{0.5ex}{\tiny$\substack{+0.6 \\ -0.4}$}\phn      &     $210.0$, $212.0$              \\
$\lambda_{\rm d}$ (deg)        & * & $270$\raisebox{0.5ex}{\tiny$\substack{+10 \\ -33}$}              &     $214$, $302$                  \\
$T^{\rm peri}_{\rm d}$ (yr)    &   & $2000$\raisebox{0.5ex}{\tiny$\substack{+21 \\ -7\phn}$}\phn      &     $1982$, $2032$                \\
$P_{\rm d}$ (yr)               &   & \phn$91$\raisebox{0.5ex}{\tiny$\substack{+18 \\ -27}$}           & \phn$56$, $186$                   \\
\enddata
\tablecomments{See Table~\ref{tbl:orvara} for details; those posteriors were adopted as input for this subsample.}
\end{deluxetable*}

\subsection{Results}
\label{A:orbit-results}

The best-fit orbit, i.e., the parameters that minimize $\chi^2$, resulted in $\chi^2=161$ for the relative astrometry alone, which is reasonable given that we fit 84 measurements with $\approx$21 free parameters (approximate because some are also constrained by RVs) and thereby 147 degrees of freedom. The subset of our new relative astrometry for planet~d independently has  $\chi^2=3.2$ for 6 degrees of freedom, implying that the astrometric errors from our analysis of VLT/ERIS, VLT/SPHERE, and JWST/NIRCam data are reasonable.

We report the resulting orbital parameters of all three planets in Table~\ref{tbl:orvara}. The marginalized posteriors (Figure~\ref{fig:full_corner}) for many of the orbital parameters are quite narrow and thus not influenced by our priors. These include all of the parameters for planet~b and all but the mean longitude at the reference epoch for planet~c ($\lambda_{\rm c} = 288$\raisebox{0.5ex}{\tiny$\substack{+23 \\ -13}$}\,deg). While uniform priors on angles like $\lambda$ are generally well-justified, the same is not true for astrophysical parameters like eccentricity and semimajor axis. The parameters that we determine to be prior-independent for $\beta$~Pic~d are the inclination $i_{\rm d}$ and longitude of ascending node $\Omega_{\rm d}$. In Table~\ref{tbl:orvara-no-X}, we report results for planet~d from the subset of orbital solutions that do not cross the orbit of planet~b.

In addition to our baseline prior, we also experimented with allowing planet~d to have some mass in our orbit fits. However, this seemed to result in attempts to `overfit' the residuals of the astrometric and radial accelerations (after removing~b and~c) by exploring unrealistically massive and eccentric orbits for planet~d. Given that we expect planet~d to have a negligible impact on the host star acceleration, these experiments justify our choice of a massless prior on planet~d (implemented as $10^{-6}\pm10^{-9}$\,\Msun in \texttt{orvara}).

\section{CMD data}
\label{app:CMD_literature}

To construct the color-magnitude diagram (CMD) shown in Figure \ref{fig:CMD} we use literature photometry and spectroscopy for a variety of sources. Planet observations for HR~8799~bcde are from \citet{2025AJ....169..209B,2026AJ....171..301B,2008Sci...322.1348M,2010Natur.468.1080M}; for \object{HIP~65426~b} they are from \citet{
2023ApJ...951L..20C,2019A&A...622A..80C}; for \object{HIP~99770~b} from \citet{
2026ApJ..1001L..26B,2023Sci...380..198C}; for \object{51~Eri~b} from \citet{2025AJ....169..209B,
2017AJ....154...10R}; for \object{AF~Lep~b} from \citet{
2025AJ....169..194B,2024ApJ...974L..11F}; for \object{$\beta$~Pic~b} from \citet{
2024AJ....168...51K,2013ApJ...776...15C}; Zhou et al. (2026, accepted) and for $\beta$~Pic~d from this work.

For brown dwarfs we plot synthetic photometry derived from observed spectra. The field-age object photometry is derived from JWST spectra published in \citet{
2024ApJ...973..107B}. To assemble known BPMG members, we used the UltracoolSheet \citep{best_2025_15802304}, which includes a synthesis of membership information derived from Banyan~$\Sigma$ \citep{2018ApJ...856...23G} and the literature, based on the approach described in \citet{2023ApJ...959...63S}. This yielded a list of 18 field ultracool dwarf systems with membership determinations from the UltracoolSheet, \citet{2018AJ....156...57D}, \citet{2021ApJS..253....7K}, and \citet{2021ApJ...911....7Z}. The CMD also shows \object{PSO~J318.5338$-$22.8603}, which was observed with JWST \citep{2025A&A...703A..79M}, as well as all BPMG members that have SPHEREx \citep{2026ApJ...999..139B} spectra in the library presented by \citet{
2026arXiv260422012G}.

\bibliography{bibliography}{}
\bibliographystyle{aasjournalv7}



\end{document}